\documentclass{ar2e}

\usepackage{color}

\def\beq{\begin{equation}}
\def\eeq{\end{equation}}
\def\beqar{\begin{eqnarray}}
\def\eeqar{\end{eqnarray}}

\def\pfrac#1#2{ \left( \frac{#1}{#2} \right) }

\def\iso#1#2{\mbox{${}^{#2}{\rm #1}$}}
\def\he#1{\iso{He}{#1}}
\def\li#1{\iso{Li}{#1}}
\def\be#1{\iso{Be}{#1}}
\def\bor#1{\iso{B}{#1}}
\def\car#1{\iso{C}{#1}}

\def\omb{\Omega_{\rm b}}
\def\nb{n_{\rm b}}

\def\zetax{\zeta_X}

\def\dx{2}
\def\dy{2}
\def\labsize{\footnotesize}
\def\ra{\rightarrow}

\def\avg#1{\langle #1 \rangle}

\def\ga{\mathrel{\raise.3ex\hbox{$>$\kern-.75em\lower1ex\hbox{$\sim$}}}}
\def\la{\mathrel{\raise.3ex\hbox{$<$\kern-.75em\lower1ex\hbox{$\sim$}}}}

\begin{document}

\input epsf.def   

\input psfig.sty

\jname{Annual Reviews of Nuclear and Particle Science}
\jyear{2011}
\jvol{}
\ARinfo{1056-8700/97/0610-00}

\title{The Primordial Lithium Problem}

\markboth{Primordial Lithium Problem}{Primordial Lithium Problem}

\author{Brian D. Fields
\affiliation{Departments of Astronomy and of Physics, University of Illinois, Urbana IL 61801 USA}}


\begin{abstract}
Big-bang nucleosynthesis (BBN) theory,
together with the precise WMAP cosmic baryon density,
makes tight predictions for
the abundances of the lightest elements.
Deuterium and \he4 measurements agree well
with expectations,
but \li7 observations
lie a factor $3-4$ below the BBN+WMAP prediction.
This $4-5 \sigma$ 
mismatch constitutes the cosmic ``lithium problem,''
with disparate solutions possible.
(1) Astrophysical systematics in the observations could exist
but are increasingly constrained.
(2) Nuclear physics experiments provide 
a wealth of well-measured cross-section data, but 
\be7 destruction could be enhanced
by unknown or poorly-measured resonances,
such as $\be7 + \he3 \rightarrow \car{10}^* \rightarrow p + \bor{9}$.
(3) Physics beyond the Standard Model can alter the \li7 abundance,
though D and \he4 must remain 
unperturbed;  we discuss such scenarios,
highlighting decaying Supersymmetric particles
and time-varying fundamental constants.
Present and planned experiments could reveal which
(if any) of these is the solution to the problem.
\end{abstract}

\maketitle

\section{Introduction}

Big bang nucleosynthesis (BBN) describes the production of
the lightest nuclides--D, \he3, \he4, and \li7--at times
$\sim 1$ sec to $\sim 3$ min after the big bang.
Theoretical predictions of the light element abundances
are well-understood and
rest on the secure microphysics of nuclear cross sections
and Standard Model weak and electromagnetic interactions
\cite{wfh,gary}.
These predictions are in broad quantitative agreement
with measured primordial light element abundances derived
from observations in the local and high-redshift universe.
This concordance represents a great success of
the hot big bang cosmology, and makes BBN our earliest
reliable probe of the universe.

BBN has dramatically changed over the past
decade in response to the cosmological revolution
sparked by the recent flood of new observations.
Notably, WMAP measurements of the
cosmic microwave background (CMB) radiation
have precisely determined the cosmological 
baryon and total matter contents \cite{wmap1,wmap7},
while high-redshift supernova observations
reveal that the universe recently entered a phase of
accelerated expansion \cite{riess98,perlmutter99,tonry03,wood-vasey07}.
These and other observations 
reveal the {\em existence} and tightly
determine the {\em abundance} of
both dark matter and dark energy 
on cosmic scales where they dominate the mass-energy
budget of the universe today.
Yet the nature of both dark matter and dark energy
remains unknown.
Thus 21st-century cosmology finds itself in
a peculiar state of  ``precision ignorance;'' 
this situation is particularly exciting for particle
physics because both 
dark matter and dark energy demand physics beyond
the Standard Model.

BBN has played a central role in the development of this
new cosmology.
CMB data now measure of the cosmic
baryon density, independently of BBN, and with
high precision.  This casts BBN in a new light:
a comparison of these two measures of cosmic baryons
provides a strong new test of the basic hot big bang framework
\cite{st}.
Moreover, this test can now be performed in a new way,
using the CMB baryon density as an {\em input} to the BBN
calculation, which outputs predictions
for each light element abundance; these can then
each be directly compared to light element observations \cite{cfo2002}.
The result is that deuterium shows spectacular agreement between
BBN+CMB predictions and high-redshift observations, and \he4
shows good agreement.  However, using the first-year WMAP data,
\li7 showed a discrepancy of a factor $2-3$,
representing a $2-3\sigma$ disagreement between observations and theory
\cite{cfo2003,cyburt,coc,cuoco}.  This disagreement has
worsened over time, now standing at a factor $3-4$ in abundance
or $4-5\sigma$:  this is the ``lithium problem.''

In this paper we present an overview of the lithium problem,
accessible to nuclear and particle physicists 
and astrophysicists.  
Broader reviews of primordial nucleosynthesis and
its relation to cosmology and particle physics
are available \cite{gary,pp2010,jp2009,iocco2009}.

In Section \ref{sect:SBBN}, we trace the origin
of the lithium problem, with a focus on the physics of 
BBN \li7 production, the nature and precision of
light element abundance measurements, and 
the state of light element concordance in view 
of the CMB-measured cosmic baryon density.
We review possible solutions to the lithium problem in
Section \ref{sect:solutions}:
(i) astrophysical systematic uncertainties in lithium
abundances and/or their interpretation; or
(ii) new or revised nuclear physics inputs to the BBN calculation,
in the form of increased mass-7 destruction via
novel reaction pathways or by resonant enhancement of
otherwise minor channels; or
(iii) new physics -- either particle processes beyond the Standard Model
occurring during or soon after BBN, or large changes to the
cosmological framework used to interpret light element (and other) 
data.
We close by summarizing the near-future outlook 
in Section \ref{sect:outlook}.

\section{Standard BBN in Light of WMAP: the Lithium Problem Emerges}

\label{sect:SBBN}

\subsection{Standard BBN Theory}

The cosmic production of light nuclides is the result of
weak and nuclear reactions in the context of an expanding,
cooling universe. 
``Standard'' BBN refers to the scenario for light element production
which marries the Standard Model of particle physics
with the ``standard'' ($\Lambda$CDM) cosmology, with:
\begin{enumerate}
\item
gravity governed by General Relativity

\item
a homogeneous and isotropic universe (cosmological principle)

\item
the microphysics of the Standard Model of particle physics

\item
the particle content of the Standard Model,
supplemented by dark matter and dark energy

\end{enumerate}
Under these assumptions, the expansion of the
universe is governed by the Freidmann equation
\beq
\label{eq:Fried}
\pfrac{\dot{a}}{a}^2 \equiv H^2 = \frac{8\pi}{3} G \rho
\eeq
where $a(t)$ is the
dimensionless cosmic scale factor (related to redshift $z$ via
$1 + z = 1/a$), and $H = \dot{a}/a$ is the 
universal expansion rate.
The total cosmic mass-energy density $\rho = \sum \rho_i$ sums
contributions from all cosmic species $i$.

By far the largest contribution to the density comes from
{\em radiation}:  relativistic species for which $m \ll T$ (with $T$
the temperature),
namely blackbody photons 
and $N_\nu = 3$ species of neutrinos and antineutrinos,
and $e^\pm$ pairs at $T \ga m_e$.
Cosmic {\em matter} consists of nonrelativistic
species with $m \gg T$: nucleons $n$ and $p$,
and $e^-$ at $T \la m_e$.
Since $\rho_{\rm rad} \gg \rho_{\rm matter}$, eq.~(\ref{eq:Fried})
shows that radiation dominates cosmic dynamics
during BBN.


BBN occurs entirely in this radiation-dominated epoch,
for which the energy density has 
$\rho \propto T^4$, where $T \propto 1/a$ (adiabatic cooling).
This, together with eq.~(\ref{eq:Fried}), gives $t \propto 1/T^2$,
or
\beq
t \approx 1 \ {\rm sec} \ \pfrac{1 \ {\rm MeV}}{T}^2
\eeq

Light-element formation depends crucially on the relative amounts
of baryons (nucleons) and radiation, 
parameterized by the baryon-to-photon ratio
\beq
\label{eq:eta}
\eta \equiv \frac{\nb}{n_\gamma} = 2.74 \times 10^{-8} \omb h^2 \ \ .
\eeq
Here $\omb = \rho_{\rm b}/\rho_{\rm crit}$ 
and $\rho_{\rm crit} = 3H_0^2/8\pi G$, with $H_0$ the present 
expansion rate (Hubble parameter).
In standard BBN, $\eta$ is the {\em only} free parameter
controlling primordial light element abundances.

Initially, cosmic baryons are in the
form of free nucleons $n$ and $p$.
For $T \ga 1$ MeV and thus $t \la 1$ sec,
weak interactions are rapid (rates per nucleon 
$\Gamma_{n \leftrightarrow p} \gg H$)
and thus 
\beqar
\label{eq:weak}
n \nu_e & \leftrightarrow &  p e^- \\
p \bar{\nu}_e & \leftrightarrow &  n e^+ 
\eeqar
drive neutron and protons to
an equilibrium ratio
\beq
\frac{n}{p} = e^{-(m_n-m_p)/T} \ \ .
\eeq
Thinking of the nucleon as a two-level system,
this simply the Boltzmann ratio of
the excited to ground state populations.

At $T = T_{\rm f} \approx 1$ MeV,  the $n-p$ interconversion (eq.~\ref{eq:weak})
stop as the weak interaction ``freezes out'' 
($\Gamma_{n \leftrightarrow p} \ll H$),
fixing $n/p \approx e^{-(m_n-m_p)/T_{\rm f}} \sim 1/6$.
Deuterium production $p(n,\gamma)d$ occurs, but is stymied by the
large number of photons per baryon $n_\gamma/\nb = 1/\eta \sim 10^9$,
which leads to effective deuteron photodissociation
by the $E_\gamma > B_{\rm d} = 2.22$ MeV tail of the Planck
distribution.
During this time,
free neutron decay reduces the neutron-to-proton ratio to
$n/p\approx 1/7$.

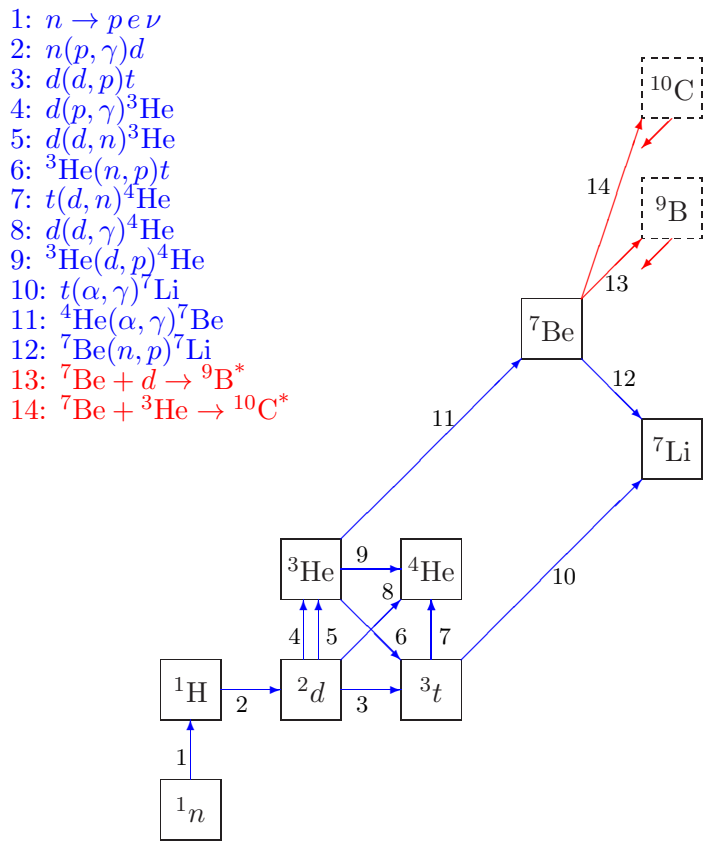
\begin{figure}

\setlength{\unitlength}{4mm}
\begin{picture}(20,28)
\linethickness{0.05mm}
\put(5,1){\framebox(\dx,\dy)[h]{${}^{1}n$}}
\put(5,5){\framebox(\dx,\dy)[h]{${}^{1}{\rm H}$}}
\put(9,5){\framebox(\dx,\dy)[h]{${}^{2}{d}$}}
\put(13,5){\framebox(\dx,\dy)[h]{${}^{3}{t}$}}
\put(9,9){\framebox(\dx,\dy)[h]{${}^{3}{\rm He}$}}
\put(13,9){\framebox(\dx,\dy)[h]{${}^{4}{\rm He}$}}
\put(21,13){\framebox(\dx,\dy)[h]{${}^{7}{\rm Li}$}}
\put(17,17){\framebox(\dx,\dy)[h]{${}^{7}{\rm Be}$}}
\put(21,21){\dashbox{0.2}(\dx,\dy)[h]{${}^{9}{\rm B}$}}
\put(21,25){\dashbox{0.2}(\dx,\dy)[h]{${}^{10}{\rm C}$}}
\color{blue}
\put(6,3){\vector(0,1){\dy}}  
\put(7,6){\vector(1,0){\dx}} 
\put(11,7){\vector(1,1){\dx}} 
\put(9.75,7){\vector(0,1){\dy}} 
\put(10.25,7){\vector(0,1){\dy}} 
\put(11,6){\vector(1,0){\dx}} 
\put(14,7){\vector(0,1){\dy}}  
\put(14,7){\vector(0,1){\dy}}  
\put(11,9){\vector(1,-1){\dy}}  
\put(11,10){\vector(1,0){\dx}}  
\put(11,11){\vector(1,1){6}}  
\put(15,7){\vector(1,1){6}}  
\put(19,17){\vector(1,-1){\dy}}  
\color{red}
\put(19,19){\vector(1,1){\dx}}  
\put(19,19){\vector(1,3){\dy}}  
\put(22,21){\vector(-1,-1){1}}  
\put(22,25){\vector(-1,-1){1}}  
\color{black}
\put(5.5,3.5){\labsize 1}
\put(7.5,5.25){\labsize 2}
\put(11.5,5.25){\labsize 3}
\put(9.25,7.5){\labsize 4}
\put(10.5,7.5){\labsize 5}
\put(12.8,7.5){\labsize 6}
\put(14.25,7.5){\labsize 7}
\put(12.35,9.){\labsize 8}
\put(11.5,10.25){\labsize 9}
\put(18.,9.5){\labsize 10}
\put(14.,14.75){\labsize 11}
\put(20.,16.1){\labsize 12}
\put(19.75,19.25){\labsize 13}
\put(19.15,22.5){\labsize 14}
\color{blue}
\put(0,28){1: $n \rightarrow p \, e \, \nu$}
\put(0,27){2: $n(p,\gamma)d$}
\put(0,26){3: $d(d,p)t$}
\put(0,25){4: $d(p,\gamma)\he3$}
\put(0,24){5: $d(d,n)\he3$}
\put(0,23){6: $\he3(n,p)t$}
\put(0,22){7: $t(d,n)\he4$}
\put(0,21){8: $d(d,\gamma)\he4$}
\put(0,20){9: $\he3(d,p)\he4$}
\put(0,19){10: $t(\alpha,\gamma)\li7$}
\put(0,18){11: $\he4(\alpha,\gamma)\be7$}
\put(0,17){12: $\be7(n,p)\li7$}
\color{red}
\put(0,16){13: $\be7+d \ra \bor{9}^*$}
\put(0,15){14: $\be7+\he3 \ra \car{10}^*$}
\end{picture}

\caption{
Simplified BBN nuclear network:  12 normally important reactions
shown in blue, and proposed/tested new reactions in red.
\label{fig:network}
}
\end{figure}

At $T \approx 0.07$ MeV, blackbody photons become ineffective
to photodissociate deuterium.  The deuteron abundance
rapidly rises, and from this all of the light elements are built
via strong (i.e., nuclear) interactions.
A simplified reaction network appears in Fig.~\ref{fig:network},
highlighting the reactions which dominate production of
the light nuclides.
In contrast to much of stellar nucleosynthesis,
for BBN the number of key reactions is small and
well-defined, and all of the important reactions have
been measured in the laboratory at the relevant energies;
no low-energy extrapolations are needed.

\begin{figure}
\psfig{figure=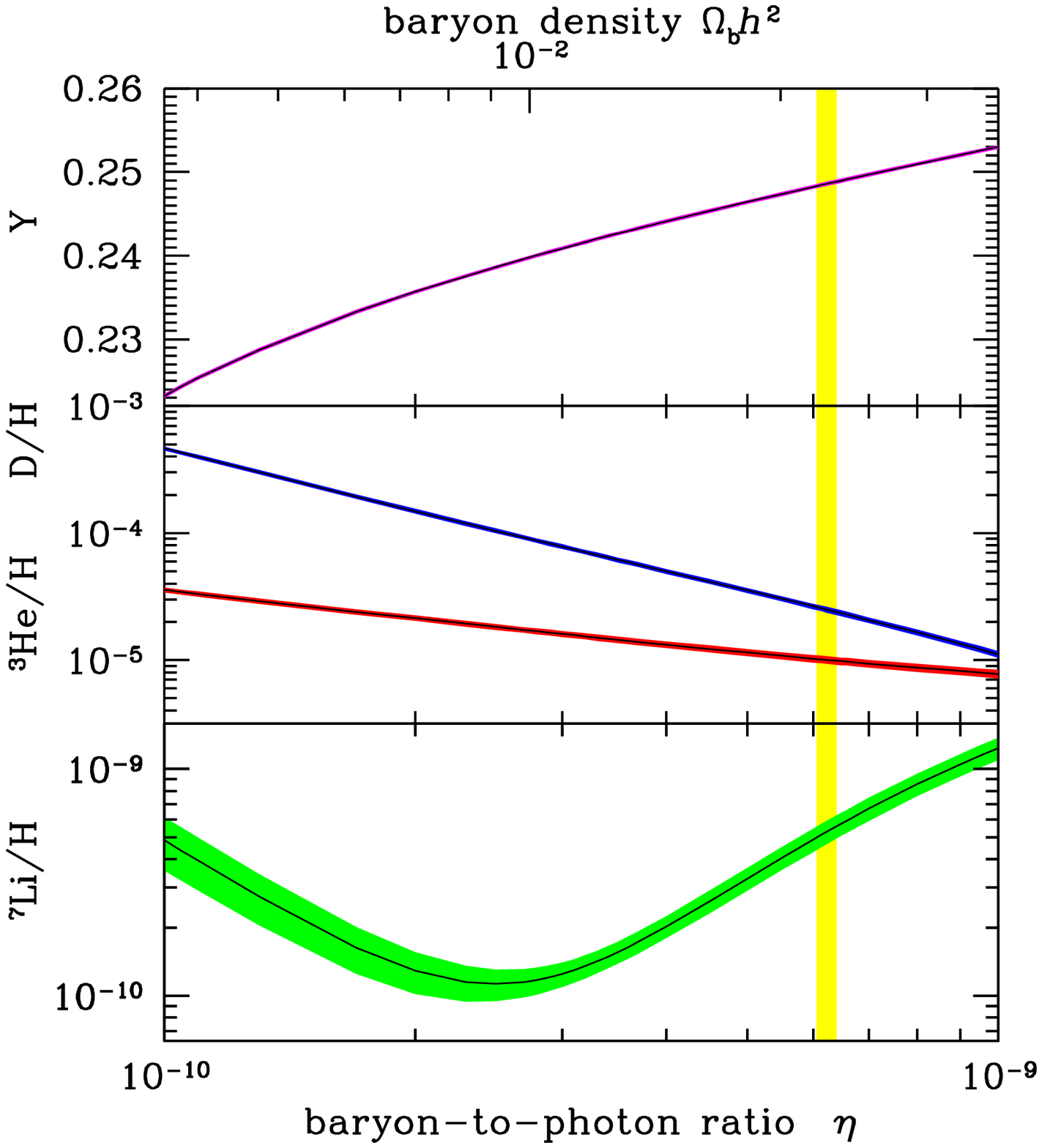,height=5in}
\caption{
BBN theory predictions for
light nuclide abundances vs baryon-to-photon ratio
$\eta$.  Curve widths: $1\sigma$ theoretical uncertainties.
Vertical band: WMAP determination of $\eta$.
}
\label{fig:schramm}
\end{figure}

Figure~\ref{fig:schramm} shows the standard BBN
light-element abundances as a function of the
single free parameter $\eta_{10} = 10^{10} \eta$ (eq.~\ref{eq:eta}).
The vertical yellow band is the WMAP $\eta$ range (see \S \ref{sect:cmb}).
We see that the \he4 abundance is weakly sensitive to
$\eta$ (note that 
the zero is suppressed in the top-panel abscissa).
In contrast, deuterium drops strongly with $\eta$
and \he3 decreases substantially.  The \li7 abundance 
is plotted after \be7 decay and thus sums both mass-7 species.
\li7 production dominates the mass-7 abundance in the low $\eta$
regime of the plot, while \be7 production dominates at the
high-$\eta$ regime, leading to the `dip'' behavior.

The envelopes around the curves in Fig.~\ref{fig:schramm} 
correspond to the $1\sigma$ uncertainties in the abundance
predictions.  These uncertainties are propagated from
the error budgets--statistical and systematic--of
the twelve dominant reactions shown in Fig.~\ref{fig:network}
\cite{kr,skm,fiorentini,hata,nb,cfo2001,coc2003,descouvemont,serpico,cyburt}.
The uncertainties in \he4 are tiny ($< 1\%$),
those in D and \he3 are small ($\sim 7\%$), while
the \li7 uncertainties are the largest ($\sim 12\%$ in the
high-$\eta$ regime of interest).

Several aspects of lithium production are noteworthy.
Mass-7 is produced both as \li7 and as \be7;
the $\be7 \stackrel{\rm \tiny EC}{\longrightarrow} \li7$ electron
capture occurs long after BBN ceases
\cite{rishi}.  
The $\he3(\alpha,\gamma)\be7$ channel
dominates \be7 production.
Destruction occurs via $\be7(n,p)\li7$ followed
by the rapid $\li7(p,\alpha)\he4$ reaction.
Finally, \li6 production in standard BBN is very small:
$\li6/{\rm H} \simeq 10^{-14}$, or $\li6/\li7 \la 10^{-4}$ 
\cite{tsof,elisa99}.

\subsection{Light Element Observations}

\label{sect:obs}

Measuring the primordial abundance of any light element
remains challenging.  The BBN levels set at $z \sim 10^{10}$
are reliably accessible only in sites at $z \le 3$ and 
oftentimes $z \sim 0$.  Other nucleosynthesis processes
have intervened, as evidenced by the nonzero metallicity
of all measured astrophysical systems.  Thus one seeks
to measure
light elements in the most metal-poor systems, 
and then to obtain {\em primordial abundances} requires
extrapolation to
zero metallicity .
Our discussion will follow closely recent
treatments in refs.~\cite{cfo2008,ceflos2009,ceflos2010}. 

\subsubsection{Deuterium, Helium-3, Helium-4} 

Deuterium can be measured directly at high redshift.
It is present in distant neutral hydrogen gas clouds which
are seen in absorption along sightlines to 
distant quasars.  At present there are 
seven systems with robust deuterium
measurements
\cite{bt98,bt98b,omeara2001,kirkman,omeara2006,pettini2008}. 
These lie around redshift $z \sim 3$ and
have metallicity $\sim 10^{-2}$ of solar; thus 
deuterium should be primordial.
For these systems,
\beq
\label{eq:deut}
\frac{\rm D}{\rm H} = (2.82 \pm 0.21) \times 10^{-5}
\eeq
where the error has been 
inflated by a the reduced $\chi^2_\nu = 2.95$.

Helium-4 can be measured in emission from nearby
metal-poor galaxies (extragalactic H {\sc ii} regions).
The challenge is to reliably infer abundances at the
needed $\la 1\%$ level.
Several recent analyses differ due to systematics
in the extraction of abundances from nebular lines
\cite{os2001,os2004,peimbert,izotov,thuan}.
The mass fraction of \cite{os2004}
\beq
Y_p = 0.249 \pm 0.009
\eeq
has the largest and most conservative
measure of the error budget; the
allowed range overlaps with analyses of other groups.

Helium-3 is at present only accessible  in our Galaxy's
interstellar medium \cite{bania}.  This unfortunately means
it cannot be measured at low metallicity, and so
its primordial abundance cannot be determined reliably
\cite{vofc}; we will not use \he3 to constrain BBN.

\subsubsection{Lithium-7}

\begin{figure}
\psfig{figure=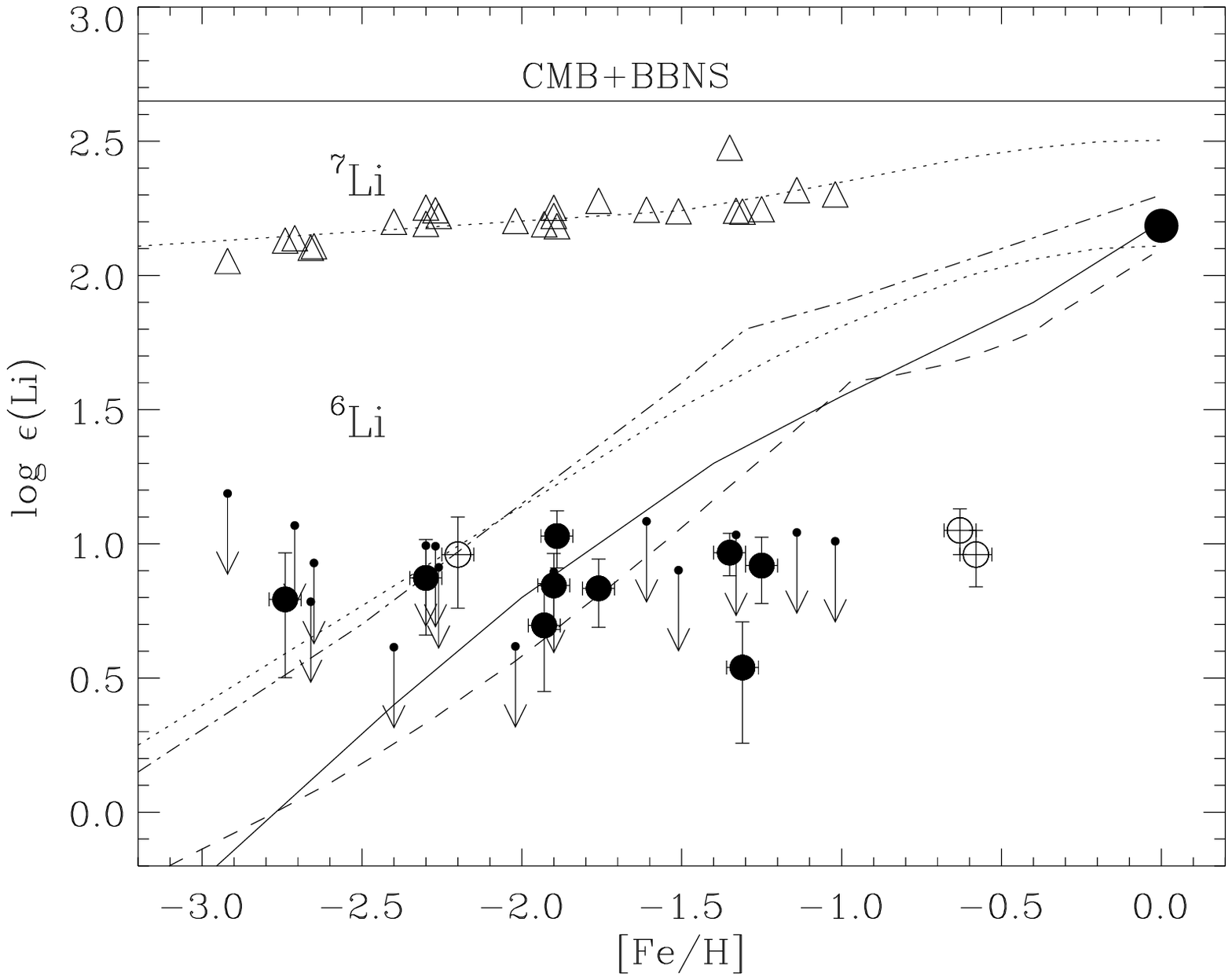,height=5in}
\caption{
Lithium abundances in selected metal-poor Galactic halo stars,
from \cite{asplund} with permission.
Fore each star, elemental ${\rm Li}=\li6+\li7$ 
is plotted at the star's metallicity
$[{\rm Fe/H}] = \log_{10}[(Fe/H)_{\rm obs}/({\rm Fe/H})_\odot]$.
The flatness of Li vs Fe is the ``Spite plateau''
and indicates that the bulk of the lithium
is unrelated to Galactic nucleosynthesis processes
and thus is primordial. 
The horizontal band gives the BBN+WMAP prediction;
the gap between this and the plateau illustrates the
\li7 problem.
Points below the plateau show \li6 abundances;
the apparent plateau constitutes the \li6 problem.
\label{fig:LiObs}
}
\end{figure}

Lithium is measured in the atmospheres of metal-poor stars
in the stellar halo (Population II) of our Galaxy.
Due to convective motions, surface material in stars
can be dragged to the hot interior where lithium
is burned readily; this effect is seen in
low lithium abundances in {\em cool} halo stars.  Fortunately,
the {\em hottest} (most massive) halo stars have thin
convection zones, and 
show no correlation between
lithium and temperature.  We consider
only lithium abundances in these stars.

Figure~\ref{fig:LiObs} shows 
lithium and iron abundances
for a sample of halo stars \cite{asplund}.  
Li/H is nearly independent of Fe/H;
this flat trend is known as the ``Spite plateau'' in honor of
its discoverers \cite{spite}.
But heavy elements such as iron (``metals'')
{\em increase} with time as Galactic nucleosynthesis proceeds
and matter cycles in and out of stars.
Thus the Spite plateau indicates that most halo star lithium is uncorrelated
with Galactic nulceosynthesis and hence, lithium is primordial.

Moreover, the Spite plateau {\em level}  measures the primordial abundance.
Thanks to a sustained effort of several groups
\cite{asplund,rnb,bonifacio,korn,sbordone,aoki,hosford09,hosford10,melendez,g-h}, 
a large sample of halo stars have measured lithium abundances.
The dominant error are systematic.
A careful attempt to account for the full lithium error budget
found \cite{rbofn}
\beq
\frac{\rm Li}{\rm H} = (1.23^{+0.68}_{-0.32}) \times 10^{-10}
\eeq
with this 95\%CL error budget dominated by systematics
(see also  \S \ref{sect:astrophys}).

Finally, it is encouraging to note that
lithium has now been
seen in stars in an accreted 
metal-poor dwarf galaxy.
The Li/H abundances are consistent with Spite plateau, pointing to its
universality
\cite{monaco}.


\subsubsection{Lithium-6}

Due to the isotope shift in atomic lines,
\li6 and \li7 are in principle distinguishable in spectra.
In practice, the isotopic splitting is several times smaller
than the thermal broadening of stellar lithium lines.
Nevertheless, the isotopic abundance remains encoded
in the detailed {\em shape} of the lithium absorption profile.

High-spectral-resolution lithium measurements in halo stars
attain the precision needed to see isotope signatures.
\li6 detections have been claimed,
in the range \cite{asplund}
\beq
\frac{\li6}{\li7} \simeq 0.05
\eeq
Fig.~\ref{fig:LiObs} shows the inferred \li6/H
abundance for some of these stars; its constancy
with metallicity is strikingly reminiscent of the
ordinary Spite plateau and similarly seems to
suggest a primordial origin.

Lithium-6 observations remain controversial.
It has been argued that stellar convective motions
can alter the delicate lineshapes and mimic \li6
\cite{cayrel2007}.
Thus there are only a few halos stars for which there is
widespread agreement that \li6 has even been detected.
Thus, the conservative approach is to take the \li6 observations
as upper limits, though it is of interest to see 
what is required to explain the ``\li6 plateau,'' if it exists.
Regardlessly, the isotopic searches confirm
that most of primordial lithium is indeed \li7.

\subsection{Microwave Background Anisotropies as a Cosmic Baryometer}

\label{sect:cmb}


It is difficult to overstate the cosmological impact 
of the stunningly precise CMB measurements by WMAP and other experiments. 
The temperature and polarization anisotropies 
encode a wealth of cosmological information.
Temperature fluctuations robustly record 
acoustic oscillations of the (re)combining baryon-photon plasma
within dark matter potential;
for a review, see \cite{hd}.
One of the most precise and robust results is the
measurement of the cosmic baryon density and thus of $\eta$.

\begin{figure}
\psfig{figure=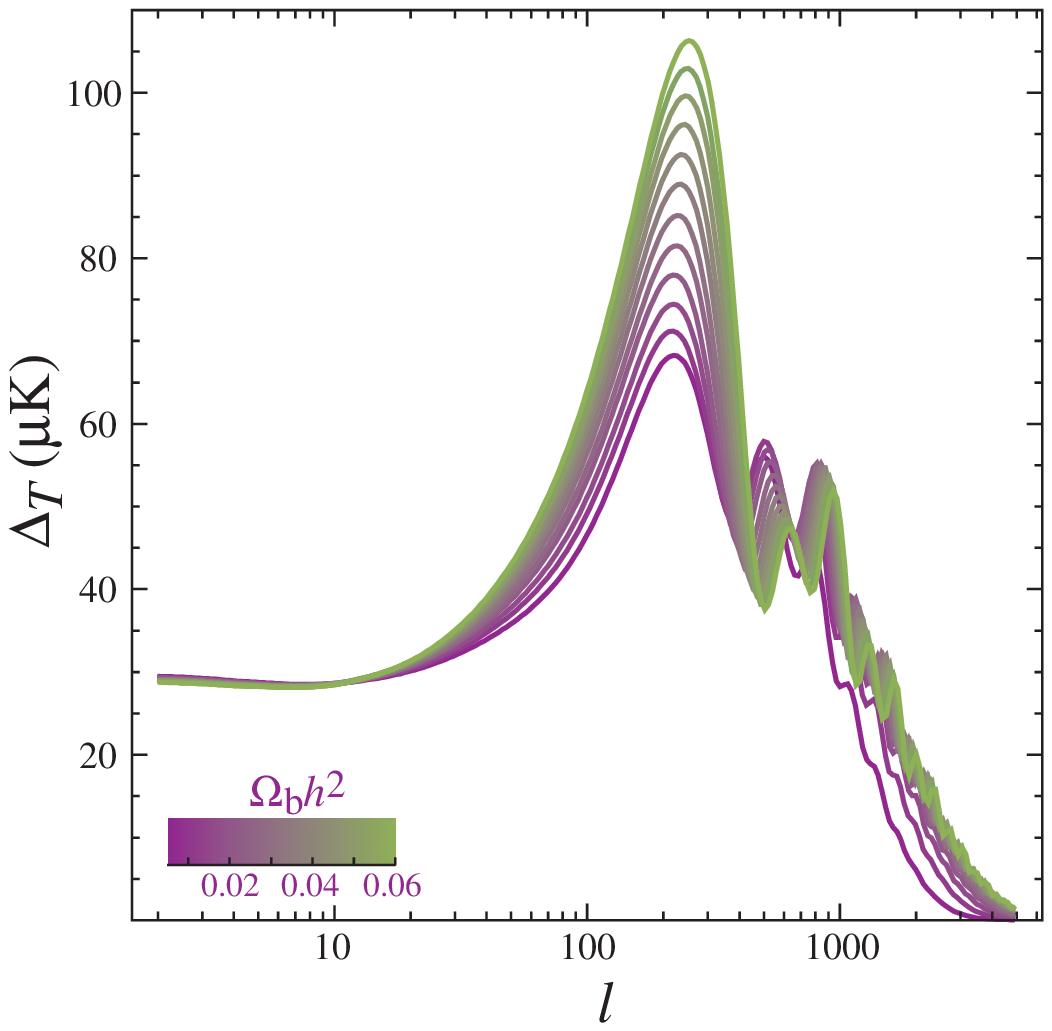,height=5in}
\caption{
CMB sensitivity to cosmic baryon content.  
Predictions for temperature anisotropy 
(rms of temperature fluctuation $\Delta_T^2 = \avg{T_\ell}^2 - \avg{T}^2$)
plotted as a function of angular scale (multipole $\ell$, roughly corresponding
to angular size $180^\circ/\ell$).
Baryon density is seen to be encoded in the values and particularly
the ratios of the peak heights.
Figure from \cite{hd}, with permission.
\label{fig:cmb}
}
\end{figure}

Figure~\ref{fig:cmb}
shows the sensitivity of the temperature anisotropy
to the baryon density, as a function of angular
scale (multipole) on the sky.
Broadly speaking, increasing baryon density 
amplifies the odd peaks and depresses the
even peaks. Accurate measurements of these
peaks by WMAP and other experiments pins down the baryon
density.  The most recent 7-year WMAP data release gives
\beq
\eta = (6.19 \pm 0.15)\times 10^{-10}
\eeq
a 2.4\% measurement!

\subsection{Assessing Standard BBN: the Lithium Problem(s) Revealed}

Prior to WMAP, BBN was the premier means of determining the
cosmic baryon density.  Standard BBN has one free parameter, $\eta$, 
but three light elements have well-measured primordial abundances:
D, \he4, and \li7.
Thus the problem is overdetermined: each element ideally selects
a given value of $\eta$, but allowing for uncertainties
actually selects a range of $\eta$.
If the different ranges are concordant, then BBN and cosmology are
judged successful, and the cosmic baryon density
is measured.  This method typically specifies $\eta$ to
within about a factor $\sim 2$ \cite{ytsso,wssof}. 

\begin{figure}
\psfig{figure=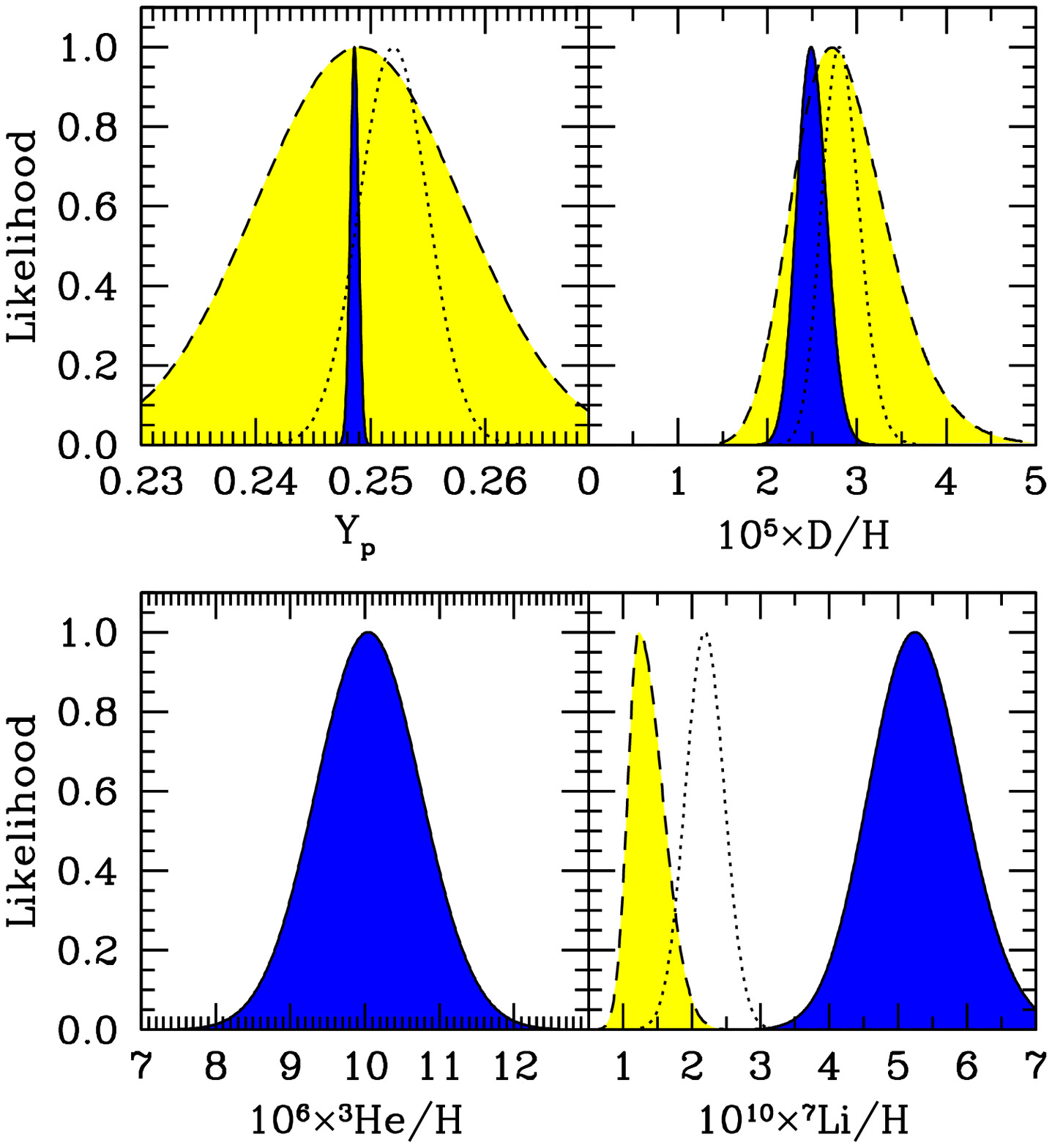,height=5in}
\caption{
Comparison of BBN+WMAP predictions and observations, from
\cite{cfo2008}.
Plotted are likelihood distributions for light nuclide abundances.
{\em Blue curves}:  theory likelihoods
predicted for standard BBN using the cosmic baryon density 
determined by WMAP\cite{wmap5}.
{\em Yellow curves}:  observational likelihoods based on
primordial abundances as in \S \ref{sect:obs}.
{\em Dotted curves}:  observational likelihoods
for different analyses of abundance data; the difference between
these and the yellow curves gives a sense of the systematic errors.
Note the spectacular agreement of D/H, and in contrast the
strong mismatch between \li7 theory and data, which 
constitutes the lithium problem.
\label{fig:likely}
}
\end{figure}

The exquisite precision of the CMB-based 
cosmic baryon density suggests a new way of 
assessing BBN \cite{st,cfo2002}. 
We exploit the CMB precision by using $\eta_{\rm mwap}$
as an {\em input} to BBN.
This removes the only free parameter in the standard theory.
Propagating errors, 
we compute likelihoods for all of the light elements.
Fig.~\ref{fig:likely}
shows these likelihoods \cite{cfo2008} based on WMAP data \cite{wmap5}.
Also shown are measured primordial abundances as discussed above.

Figure~\ref{fig:likely} shows that deuterium observations 
are in spectacular agreement with predictions--the 
likelihoods literally fall on top of each other.
This concordance links $z\sim 3$ abundance measurement with
$z\sim 10^{10}$ theory and $z \sim 1000$ CMB data,
and represents a triumph of the hot big bang cosmology.
We also see that \he4 predictions are in good agreement
with observations.  
And, as noted in \S \ref{sect:obs}, no reliable primordial \he3 measurements
exist.

Turning to \li7,  the BBN+WMAP predictions
and the measured primordial abundance completely disagree:
the predictions are substantially {\em higher} than the observations.
Depending on the treatment of systematic errors
in the measured Li/H,
the discrepancy is 
a factor $\rm Li_{bbn+wmap}/Li_{\rm obs} = 2.4-4.3$,
representing a $4.2-5.3 \sigma$ discrepancy.
This substantial mismatch constitutes the 
{\em lithium problem} (i.e., the \li7 problem).

Finally, as noted in \S \ref{sect:SBBN},
standard BBN predicts an unobservable \li6/H abundance
and \li6/\li7 ratio far below the putative \li6 plateau.
To the extent that the \li6 plateau is real, this would constitute
a second Li problem--the \li6 problem.

Hereafter we will focus largely on the well-established
\li7 problem,  but where appropriate we will mention 
the \li6 problem.

\section{Solutions to the Lithium Problem(s)}

\label{sect:solutions}

We have seen that the lithium problem was brought into sharp relief
with the advent of the WMAP era, has become increasingly
acute since.  Possible solutions fall in three broad classes
corresponding which part of the preceding analysis
is called into question:
\begin{enumerate}

\item
{\em Astrophysical} solutions revise the measured primordial
lithium abundance.

\item
{\em Nuclear Physics} solutions alter the reaction flow into
and out of mass-7.

\item
Solutions {\em beyond the Standard Model}
invoke new particle physics or nonstandard cosmological physics.

\end{enumerate}
We consider each in turn.

\subsection{Astrophysical Solutions}

\label{sect:astrophys}

We first consider the possibility that BBN predictions are sound,
i.e., standard cosmology and particle physics are correct,
and the nuclear physics of mass-7 production is properly 
calculated.  If so, then the 
measured value of the primordial lithium abundance
must be in error.  In particular, the true value must be
higher by a factor 3--4.

As described in \S \ref{sect:obs},
lithium abundances are measured via absorption lines in
the photospheres of primitive, low-metallicity stars.
For each star, the lithium
line strength is used to infer the Li/H abundance.
Lithium abundances are nearly insensitive to metallicity
(Fig.~\ref{fig:LiObs})--this Spite plateau
implies that lithium is independent of Galactic nucleosynthesis
and is primordial, and the plateau level
is taken as the primordial abundance.
If missteps exist in this chain of reasoning,
the lithium problem could potentially be alleviated.

For example, systematic errors could shift
Li/H ratios in each star.  
We seek
the {\em total} lithium content of the stellar photosphere, i.e.,
summed over all ionization states.
However, the single (!) accessible 670.8 nm lithium line 
is sensitive only to neutral ${\rm Li}^0$.
But in the stars of interest, lithium
is mostly singly ionized. One must therefore introduce
a large ionization correction ${\rm Li}^+/{\rm Li}^0$
that is exponentially sensitive to the stellar 
temperature.
Thus, a systematic shift upward in the temperature scale for halo
stars would increase all stellar Li abundances
and raise the Spite plateau towards
the BBN+WMAP prediction.

In practice, accurate determination of stellar temperatures 
remains non-trivial, 
because the emergent radiation is not a perfect Planck curve
(else no lithium lines!), nor is local thermodynamic
equilibrium completely attained in the stellar atmospheres.  
Quantitatively, the needed systematic shift
in the temperature scale is about $\Delta T_{\rm eff} \simeq 500-600$ K
upward, a $\sim 10\%$ increase over fiducial values
(outside of previous claimed errorbars) \cite{fov}.
A re-evaluation of one method to determine stellar temperatures
indeed corrects the scale, typically by $\sim +200$ K
 \cite{mr}.
However, later detailed studies of the stellar temperature scale
are in good agreement with the fiducial temperature scale,
leaving the lithium problem unresolved 
\cite{hosford10,casagrande}.

An entirely separate question remains as to whether 
a star's {\em present}  lithium
content reflects its {\em initial} abundance in the star.
If the halo stars have destroyed some of their lithium, 
their present Li/H ratio sets
a {\em lower limit} to 
the primordial lithium abundance.
Indeed, given the low nuclear binding of \li7, it 
need only be exposed to relatively low stellar temperatures
($T \ga 2.5 \times 10^6$ K) to suffer substantial destruction
over the many-Gyr lifespan of a halo star.

Lithium depletion is a major diagnostic of stellar structure
and evolution \cite{wallerstein,spites,gustafsson,pinsonneault}. 
For stars of {\em solar} composition,
lithium destruction \cite{bodenheimer,iben}
\cite{iben} has long been studied
in stellar evolution models.
The major effect is convection,
which circulates photospheric material deep into the interior
(though still far from the stellar core)
where nuclear burning can occur.
Models for the evolution of 
{\em low-metallicity} stars, appropriate for the Spite plateau,
now include numerous mixing effects which can change the photospheric
lithium:  convective motions, turbulence,
rotational circulation, diffusion
and gravitational settling,
and internal gravity waves
\cite{pinsonneault,pdd,charbonnel,richard}.
These effects must occur at some level,
and models have some success in fitting some observed trends in 
halo stars.  
There is general agreement that for stars with low metallicities,
convective zones are substantially shallower than in solar-metallicity
stars, and so depletion such be much smaller than the
factor $\sim 10^2$ in the Sun \cite{gr}.
However, it remains difficult for models to quantitatively
fit all of the data \cite{bonifacio}. 

Thus {\em observational} efforts to find clues for lithium depletion
in the Spite plateau stellar data themselves remain of utmost importance.
One study found the Spite plateau in field halo stars to 
be very thin, with no detectable star-to-star variations around
the Li-Fe trend, which showed a small positive slope in Li/H vs Fe/H
\cite{rnb}.
A small lithium increase with metallicity is {\em required}
due to contamination from Galactic cosmic-ray production of 
\li7 and \li6 \cite{rbofn}.
An analysis of lithium and iron abundances in
stars from the same globular cluster
found trends consistent with lithium depletion via diffusion and 
turbulent mixing; some models suggest these effects could
remove the lithium problem entirely \cite{korn}. 
However, systematic differences 
between globular cluster and
field star lithium abundances raise concern about
globular clusters at sites for constraining primordial lithium  \cite{g-h}.

Recently, several groups have found
that at {\em very} low metallicity, $[{\rm Fe/H}] \la -3$, 
lithium abundances 
on average fall {\em below} the Spite plateau
(i.e., below the levels
seen at metallicities $-3 \la [{\rm Fe/H}] \la -1.5$)
\cite{aoki,hosford09,sbordone,melendez}.
These groups also find that the
star-to-star {\em scatter} in Li/H becomes significant
below $[{\rm Fe/H}] \la -3$. 
This appears to confirm the presence of significant lithium depletion
in at least some halo stars.

The recent evidence for lithium depletion at very low metallicities
is a major development, yet its implications for
primordial lithium remain unclear.
No significant scatter is detected in plateau stars with
$-3 \la [{\rm Fe/H}] \la -1.5$.
Also, in no stars is Li/H seen {\em above} the
plateau, and in no metal-poor stars is Li/H near the WMAP+BBN value.
Finally, while \li6 measurements are difficult and controversial,
there is general agreement that \li6 is present in at least some 
plateau stars.  This much more 
fragile isotope strongly constrains thermonuclear burning processes--if 
stellar material is exposed to
temperatures hot enough to significantly reduce \li7, \li6
should be completely destroyed.
\cite{bs}.

To summarize, determination of the primordial lithium abundance
continues to be the focus of rapid progress.  At present, however,
the observational status of primordial lithium remains
unsettled.  A purely astrophysical solution
to the lithium problem remains possible.  On the other hand,
the observed lithium trends--particularly the small lithium
scatter in temperature and metallicity, and the presence of \li6--strongly
constrain (but do not rule out) solutions of this kind.
Consequently, it is entirely possible that the lithium problem
{\em cannot} be resolved astrophysically, and thus we are
driven to seek other explanations of the discrepancy; we now turn to these.

\subsection{Nuclear Physics Solutions}

We now consider the possibility that the measured primordial
lithium abundance is correct, and the Standard Model of particle
physics and the standard cosmology are also sound.  
In this case, the lithium problem must point to 
errors in the BBN light element predictions, in the form of
incorrect implementation of standard 
cosmological and/or Standard Model physics.

However, 
the standard BBN calculation is very robust and thus difficult to perturb.
As summarized in \S \ref{sect:SBBN}, standard BBN
rests on very well-determined physics
applied in a very simple system.
The cosmological framework of BBN is that of a very
homogeneous universe (guaranteed by the smallness of the
observed CMB temperature fluctuations \cite{wmap1}),
with a cosmic expansion governed by exact expressions
in General Relativity. 
The microphysics is that of the Standard Model, also very well-determined.
The relativistic species, which comprise cosmic radiation
that dominates the energy density, are very well thermalized and
thus their properties are that of Bose-Einstein and Fermi-Dirac
gasses, for which exact expressions are also available.

The weak and strong (i.e., nuclear) interactions are also 
well-grounded in theory and calibrated empirically, but
for BBN the needed physics is complicated (nuclear networks are
large) and lies the farthest from first principles.
Thus, these are the only possible ``weak links'' in the standard BBN
calculation, and it is here that solutions to the lithium problem
have been sought.

\subsubsection{New and Revised Reactions}

One possibility is that weak and nuclear reactions in the BBN calculations
are miscalculated due to reactions that are entirely missing, or that are
included but whose rates are incorrect.
But as described in \S \ref{sect:SBBN} and seen in Fig.~\ref{fig:network},
only a relatively small number of reactions have been found to be important for
producing the light elements, and all of these have been measured in
the laboratory at the relevant energies.  Their uncertainties have also been
calculated and propagated through the BBN code, and are folded into the likelihoods
appearing in Fig.~\ref{fig:likely}.
Moreover, BBN calculations use a much more extended reaction network than
the simplified view of Fig.~\ref{fig:network},
with all initial state pairings of $A \le 7$ species present
but most practice unimportant \cite{mf,tsommf,cfo2001,iocco2007,vcc,boyd,chfo}.
Thus, to change the primordial lithium predictions requires surprises of
some kind--either (i) the cross sections
the known important reactions have uncertainties far beyond the quoted errors, or
(ii) the cross sections for normally unimportant reactions have been vastly 
underestimated.

For the important reactions seen in Fig.~\ref{fig:network}, mass-7 production
is dominated by the single reaction $\he3(\alpha,\gamma)\be7$.
While the quoted error budget in the measured cross section
is small, $\sim 7$\% \cite{cd}, absolute cross sections are difficult to measure.
However, this reaction is also crucial in the production of solar neutrinos. 
To fix the cosmic lithium problem, the $\he3(\alpha,\gamma)\be7$ normalization 
would need to be low by a factor $3-4$; if this were the case,
the \be7 and \bor{8} solar neutrino fluxes would be lower by a similar factor.
Thus we can view the sun as a reactor which probes the
$\he3(\alpha,\gamma)\be7$ rate, and the spectacular and precise agreement
between solar neutrino predictions and observations becomes a measurement of
the rate normalization which confirms the experimental results and 
removes this as a solution to the lithium problem \cite{cfo2004}.

Weak rates in BBN have received a great deal of attention over the years
\cite{dicus,seckel,dolgov,kernan,dodelson,hannestad,lopez,esposito,sf}.
The basic $n \leftrightarrow p$ interchange rates (eq.~\ref{eq:weak}) are most accurately
normalized to the neutron lifetime.  Corrections to the tree-level rates
have $\la 1\%$ effects on abundances, and thus are far too small
to impact the lithium problem \cite{sf}.

Corrections to the standard thermonuclear rates have been considered as well.
The effects of nonthermal daughter particles has been studied and found
to be negligible
\cite{vkk,boyd}.
Plasma effects, and electron Coulomb screening
are also unimportant \cite{itoh}.

Turning then to (normally) subdominant reactions, Angulo et al.~\cite{angulo} noted that 
the (nonresonant) cross section for $\be7(d,\alpha)\alpha p$ 
was poorly determined and could solve the lithium problem if it were
a factor $\sim 100$ larger.  They measured the cross section
at BBN energies, but found values a factor $\sim 10$ {\em smaller}
than had been used.

Finally, the possibility of entirely new reactions has been recently studied
by Boyd et al.~\cite{boyd}.
These authors systematically considered a large set of reactions,
some of which have been neglected in prior calculations.
The focus of this study is almost exclusively on nonresonant reactions,
with the result that even when allowing for extremely large systematic
uncertainties in known cross sections, most new channels remain unimportant.
The loophole to this analysis is the presence of new or poorly measured
resonances.

\subsubsection{Resonances}

Both standard and nonstandard reaction pathways to primordial mass-7 
are firmly anchored to experimental data.
The only remaining new nuclear physics can only intervene via
resonances which have evaded experimental detection or
whose effects have been underestimated.
Cyburt and Pospelov \cite{cp}
point out that the production of the known resonance
$\be7 + d \rightarrow \bor{9}^*(16.71 \ {\rm MeV})$
is poorly constrained experimentally;
this reaction appears in Fig.~\ref{fig:network}.
Within current uncertainties, this resonance could 
promote the
$\be7+d$ channel to become the
dominant \be7 destruction mode, and 
solving the lithium problem in an elegant manner.

Generalizing the Cyburt and Pospelov \cite{cp} suggestion,
ref.~\cite{chakraborty} searched the entire resonance
solution space for BBN.
These authors systematically considered all compound states created in
mass-7 destruction, via all possible 2-body
reactions of the form $(n,p,d,t,\he3,\he4)+(\li7,\be7)$.
Most possibilities were found to be unimportant.
However, in addition to the 
$\be7 + d \rightarrow \bor{9}^*(16.71 \ {\rm MeV})$ resonance,
two other potentially important states were identified.
The $\be7 + t \rightarrow \bor{10}^*(18.80 \ {\rm MeV})$ resonance
is known and within present uncertainties could be significant.
On the other hand, there is little data on high-lying states of $\car{10}$,
but if a $\car{10}^*(15.0 \ {\rm MeV})$ exists and has $J^\pi = 1^-$ or $2^-$,
this also could bring cosmic lithium into concordance if the reaction widths
are large enough.
This last possibility would be a homage to 
Fred Hoyle's celebrated prediction of the
$\iso{C}{12}^*(7.65)$ state which solved 
the ``carbon problem'' of stellar nucleosynthesis \cite{hoyle}.

Fortunately all of these states are experimentally accessible. 
To identify or exclude them marks the endgame for
a nuclear solution to the lithium problem.

\subsection{Solutions Beyond the Standard Model}

Finally, we turn to the most radical class of solutions
the lithium problem.  Namely, we assume that primordial lithium has been correctly measured,
and that the nuclear physics of BBN has been calculated correctly and holds no surprises.
In this case, we are forced to question the assumptions underlying the standard BBN
calculation, i.e,. we must go beyond the Standard Model of particle physics
and/or the standard cosmology.
For details beyond the overview below, see
refs.~\cite{jp2009,pp2010}.

\subsubsection{Dark Matter Decay and Supersymmetry}

The existence of dark matter is now well-established, and its cosmic abundance has been inferred
precisely; for recent reviews, see refs.~\cite{feng,hooper,gaitskell}.
If dark matter takes the form of a relic particle created in the
very early universe, it must be nonbaryonic.  
No Standard Model particles have the right properties, and
thus dark matter {\em demands} physics beyond the Standard Model.
Dark matter must of course be present during BBN, but 
ordinarily is assumed to be both
nonrelativistic and weakly interacting,
hence
unimportant to cosmic dynamics and microphysical interactions.
Similarly, dark energy is assumed to be negligible.

While the identity of the dark matter is unknown,
a simple, popular, and physically well-motivated possibility is that
dark matter today consists of relic weakly interacting massive particles
(WIMPs). In particular, if the universe begins with equal abundances
of WIMPs  and anti-WIMPs (if the two are distinct), then their abundance today is determined
by the freezeout of their annihilations.
Famously, to reproduce $\Omega_{\rm m} \simeq 0.3$ today,
the annihilations must occur at $\sim \rm TeV$ scales (well before BBN).
By happy coincidence, this is also the scale of the weak interaction,
and of current accelerator experiments--this is the ``WIMP miracle''
\cite{feng,hooper,gaitskell}.

Moreover, it is likely that
WIMPs today are the stable endpoints of a decay cascade.
If so, then the WIMPs are the daughters born in the decays of the next-lightest
particles in the cascade.  The nature of these decays is model-dependent, 
but in general produce Standard Model particles
which interact with the background plasma.
If the decays occur during or after BBN,
the interactions can change light element abundances
\cite{ekn84,lindley85,ens85,jss85,ks87,reno88,scherrer88,dehs89,eglns92,moroi93,jedamzik2006,cefo,kkm2005,
pospelov2007},
and thus potentially could solve the lithium problem(s)
\cite{jed2004li6,jedamzik2004,jcrr2006,kusakabe2007,pps2008,jedamzik2008,bjm2009,ceflos2010}.

We thus consider the effects of decaying massive particles
during or after BBN.  To get a feel for the basic physics,
consider a particle $X$ (and $\bar{X}$ if they are distinct)
with mass $m_X \gg m_p$,
which decays with lifetime $\tau_X$.
The decays can have electromagnetic 
and/or hadronic channels, and
given the massive nature of $X$, these decay products will
be very relativistic and thus nonthermal.

As these electromagnetic and/or hadronic cascades thermalize
in the cosmic environment,
they interact with the light elements, largely
via fragmentation (photodissociation or spallation).
For example, high energy nucleons $N$ fragment
\he4:
\beq
\label{eq:erode}
N + \he4 \rightarrow 
  \left\{
    \begin{array}{l}
       2n + 2p + N \\
       d + d + N \\
       d + {}^3{A} \\
       \cdots
    \end{array}
  \right.
\eeq
This reduces the \he4 abundance, but more importantly
creates new deuterium and \he3.
Furthermore, the secondary particles in eq.~(\ref{eq:erode})
are themselves nonthermal, and can initiate further
interactions with the background thermal light nuclides.
Of particular importance is the
conversion of
\be7 into \li7 with secondary neutrons
\beq
\label{eq:secn}
n + \be7 \rightarrow p + \li7 \ \ ,
\eeq
which substantially enhances mass-7 destruction because of
the lower Coulomb barrier for \li7.
Also of interest is the nonthermal
production of \li6 via secondary nonthermal deuterons
\beq
d + \he4 \rightarrow \li6 + \gamma \ \ .
\eeq
Clearly, both of these processes are of great interest
for both lithium problems.

\begin{figure}
\psfig{figure=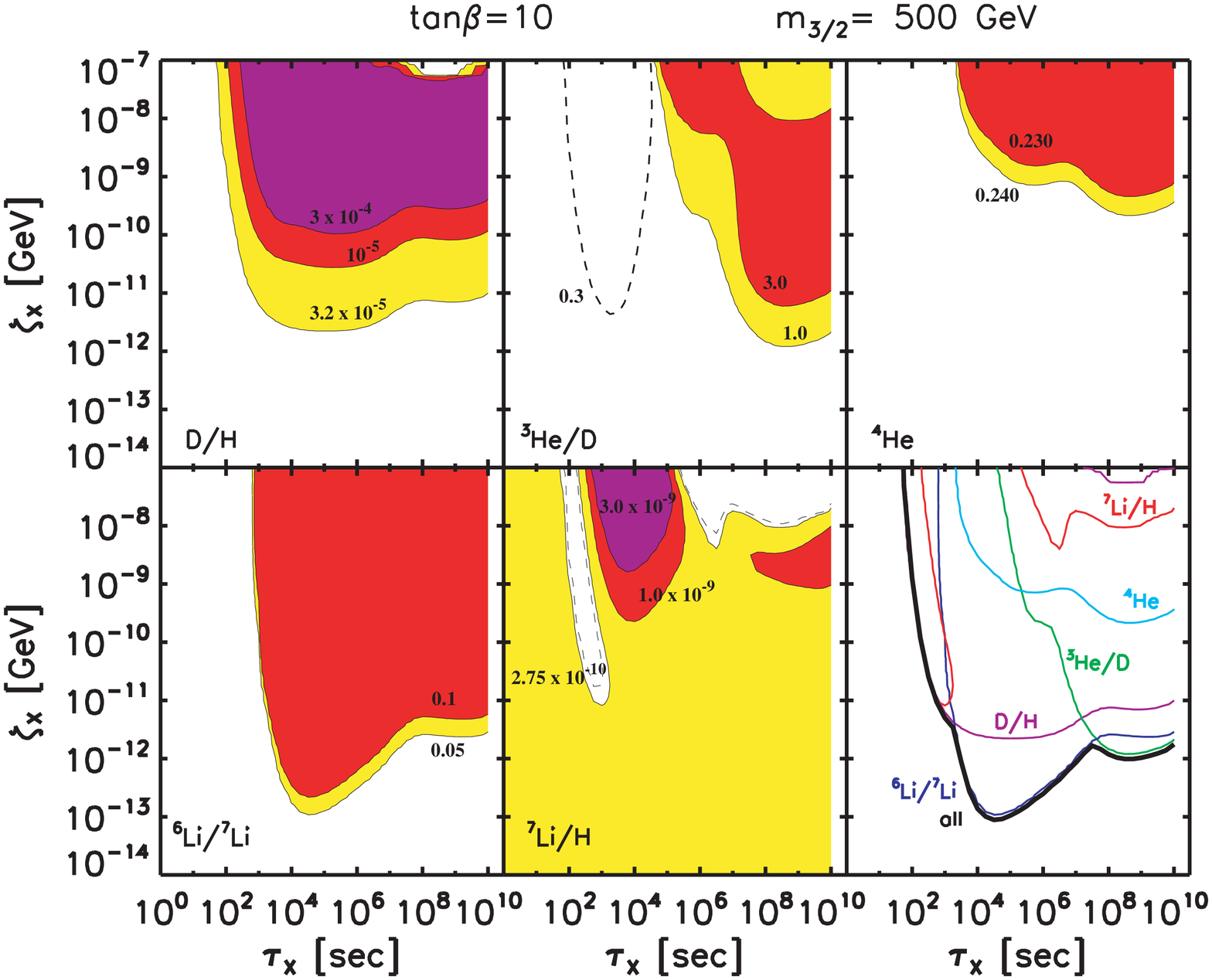,width=5.0in}
\caption{
Effect of nonthermal particle injection on light element abundances,
from \cite{ceflos2009}.
Each panel shows abundance contours in the presence of the hadronic decay
of a (neutral) particle $X$, 
plotted as a function of $X$ abundance $\zetax$ (eq.~\ref{eq:zeta})
and mean life $\tau_X$.
Results are shown for a hadronic branching ratio $B_h = 1$.
As summarized in the last panel, the parameter regions where the \li7 problem
is solved also lead to deuterium production, placing the two in tension. 
\label{fig:zeta-tau}
}
\end{figure}

\begin{figure}
\psfig{figure=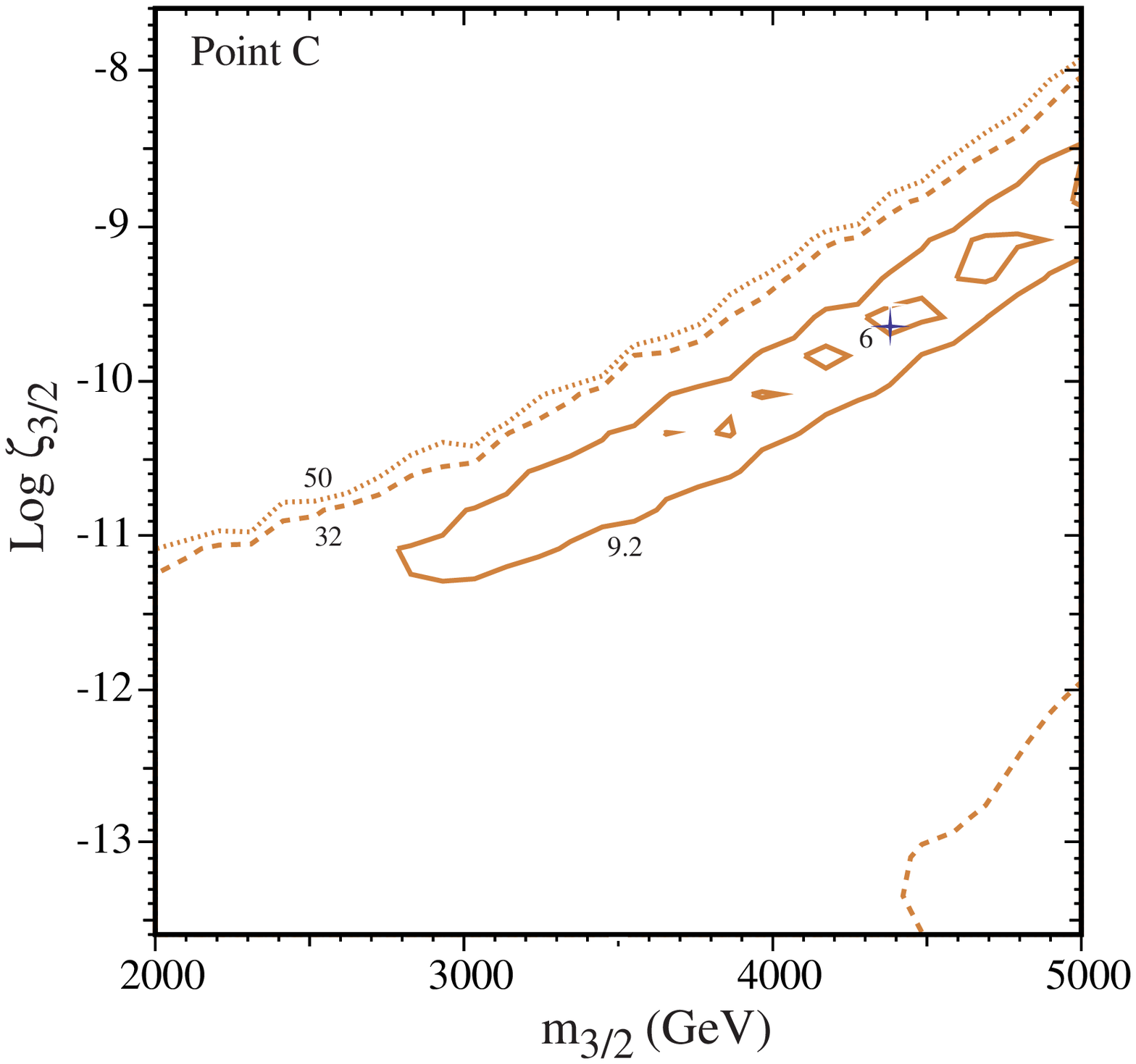,width=0.9\textwidth}
\caption{
Light element constraints on gravitino decays in the context of 
the constrained minimal 
supersymmetric Standard Model, from \cite{ceflos2010}. 
Shown are  $\chi^2$ contours based on light element abundance constraints 
and 1 effective degrees of freedom, 
over the space of gravitino mass $m_{3/2}$ and abundance 
$\zeta_{3/2}$ (eq.~\ref{eq:zeta}). 
The cross indicates the maximum $\chi^2$. 
diagonal ``archipelago'' shows the region where the lithium
problem is greatly reduced by trading off \li7 destruction for
some degree of deuterium production.
\label{fig:pointC}
}
\end{figure}

Figure \ref{fig:zeta-tau} shows light-element abundances in the presence of
a hadronically decaying particle $X$.
Abundance contours are plotted as a function of  
decay lifetime $\tau_X$, and the pre-decay
$X$ abundance 
\beq
\label{eq:zeta}
\zetax \equiv \frac{m_X n_X}{n_{\gamma}} = m_X \frac{n_X}{\nb} \eta
\eeq
In the absence of $X$ decays, cosmic expansion dilutes
baryons and $X$ particles in the same
way, and thus $\zetax$ remains constant.
Figure \ref{fig:zeta-tau} is for a hadronic branching fraction $B_h=1$, i.e.,
all decays produce hadrons.

In each panel of Fig.~\ref{fig:zeta-tau}, the colored areas indicate parameter
regions where the predicted light element abundances disagree with observations,
while the remaining white regions are allowed.
At fixed $\tau_X$, the light element perturbation is proportional to
the $X$ abundance. The limit $\zetax \rightarrow 0$ lies at the
bottom of the plot, and corresponds to the
unperturbed (standard BBN) case.
At fixed abundance $\zetax$, the light element
effects strongly depend on particle lifetime, which determines when
the light elements are perturbed.  For example, at $\tau_X \la 10$ sec, decays
occur before the light elements are formed, and so the light elements
are unaffected.   Hadronic decays dominate the perturbations in the $\tau_X \la 10^6$ sec
of most interest here;
electromagnetic decays dominate at longer times.

In the \li7/H panel, the colored
regions include the unperturbed low-$\zetax$ regime, reflecting the lithium problem
for standard BBN.  However, for a relatively narrow region with lifetimes
$\tau_X \sim 10^2-10^3$ sec, that 
the \li7 abundance is reduced and brought into accord with observation.
This arises due to the production of secondary neutrons (eq.~\ref{eq:erode}),
which facilitate mass-7 destruction via \be7-to-\li7 conversion (eq.~\ref{eq:secn});
at longer lifetimes the secondary neutrons decay.
For $10^2-10^3$ sec lifetimes to be viable, however, the other light elements
must remain in concordance.  As seen in the final, summary panel,
in the regime in which \li7 is reduced, all other constraints are satisfied
{\em except} D/H, which is unacceptably high due to secondary deuteron production.
This tension between deuterium and \li7 is a fundamental feature of
decay scenarios, and as we see may allow for solutions
but requires fine tuning.

Supersymmetry provides well-motivated candidates for decaying dark matter
\cite{peskin,olive,martin}. 
Supersymmetry doubles the particle content of nature by requiring
opposite-statistics (fermion $\leftrightarrow$ boson) partners
for every known particle. These partners are produced
copiously in the very early universe.
The lightest supersymmetric partner (LSP) 
is the stable end product of the decays of higher-mass supersymmetric particles,
and naturally becomes
a dark matter candidate
\cite{ehnos,eoss}.
These decays are a fundamental aspect of supersymmetric dark matter, 
and thus supersymmetric scenarios {\em demand} that particle decays
occur. Moreover, even in the simplest scenarios--i.e., the constrained
minimal supersymmetric Standard Model--the lifetime of the 
next-to-lightest supersymmetric partner can be long:  $\ga 100$ sec.
Thus, minimal supersymmetry holds the tantalizing possibility of
solving the lithium problem.

Figure \ref{fig:pointC} illustrates the interplay between supersymmetry
and the lithium problem \cite{ceflos2010}.  
The context is minimal supersymmetry in which the spin-3/2 gravitino
is the next-to-lightest partner, which decays into the
LSP neutralino that comprises the dark matter today.
For a benchmark scenario in this model, the superpartner mass spectrum,
lifetimes, and decay products are calculated.  These are used to compute
light element abundances, with nuclear uncertainties propagated;
the results are compared to observations.
The resulting $\chi^2$ is plotted as a function of gravitino mass $m_{3/2}$
and pre-decay abundance $\zeta_{3/2}$.
At low $\zeta_{3/2}$, we recover the unperturbed, standard  case,
where $\chi^2 \approx 32$ for one effective degree of freedom
(3 light elements $-$ 2 parameters), a poor fit.  
At high $\zeta_{3/2}$, the light element perturbations are {\em worsened}
over the standard case, and these supersymmetric parameters are
excluded by BBN.  Most interestingly, along the diagonal loop, the fit
improves, and in the interior ``islands,'' $\chi^2$ drops below 6,
corresponding to $\sim 2.4 \sigma$.  This region shows substantial
improvement over standard BBN, and physically arises as the regime 
of optimal tradeoff between \li7 destruction and deuterium production.
This region thus tantalizingly stands as a possible supersymmetric solution
to the lithium problem, albeit statistically marginal and fine-tuned.

Finally, in very recent years an entirely new aspect of decaying dark matter
alteration of BBN has emerged.
If the decaying particles are electrically charged, then 
the negatively charged dark matter can form bound states
with charged nuclei, such as $(pX^-)$, $(\he4X^-)$, and $(\be7X^-)$.
As pointed out by Pospelov \cite{pospelov2007},
these bounds states unleash a rich array of new effects,
in addition to the perturbations accompanying $X^-$ decays.
For the heavy $X^-$ of interest, the binding energy
of $(A^ZX^-)$ is
$B = Z^2 A \alpha^2 m_p/2 \approx 30 Z^2 A$ keV, 
comparable to nuclear binding.
The Bohr radius is $a = (2 \alpha Z A m_p)^{-1} \approx 1 A^{-1} Z^{-1}$ fm,
comparable to nuclear sizes. 
Bound-state Coulomb barriers are thus reduced
and $(\be7X^-$) becomes easier to destroy.

Beyond these basic effects, a new ``bound state chemistry'' can occur, whose
nature is still under active study.  One important effect \cite{pospelov2007}
is catalysis, particularly $d+(\he4X) \rightarrow \li6 + X$ 
which enhances \li6 production far above the ordinarily
small radiative capture $d(\alpha,\gamma)\li6$.
This can enhance \li6 production by orders of magnitude,
possibly addressing the \li6 problem if real, but oftentimes
overproducing \li6.  Rates for catalyzed reactions
have recently been calculated to high precision \cite{kamimura}.
Intriguingly, $(\be8X^-)$ states have a binding energy very close 
to the energy for $\be8 \rightarrow \alpha \alpha$ breakup;
if the binding energy is larger, then 
$(\be8X^-)$ is stable and can allow for cosmological production of 
\be9 \cite{pps2008,kusakabe2010}.

The general properties of $X^-$ recombination and bound states,
and their impact on BBN, have been studied in 
refs.~\cite{pospelov2007,bird2008,jedamzik2008,kusakabe2008,pps2008,pp2010}.
Looking at bound state effect only (i.e., ignoring 
decay effects), 
catalyzed \be7 destruction is effective and
catalyzed \li6 is not {\em over}produced,
for sufficiently large 
$X^-$ abundance in the regime
$\tau_X \sim 2000$ sec
\cite{bird2008}.
Indeed, there exist regions of parameter space wherein
{\em both} the \li7 and \li7 problems are solved.
Full calculation of bound state and decay effects together
is required to verify if solutions remain in specific detailed
models;
early calculations confirm solution space
exists around $\tau_X \sim 10^3$
\cite{cefos,jedamzik2008}.

Catalyzed production of \be9 also occurs and
is constrained by the non-observation of a beryllium
``Spite plateau;'' the resulting limits
on $\tau_X$ are comparable to those imposed by
limits on primordial \li6 \cite{pps2008}.
\be9 constraints have the added advantage that
they rely on elemental abundances which are much 
simpler to obtain reliably than isotopic abundances.
However, very recently \be9 was investigated using updated
catalysis rates from ref.~\cite{kamimura}, which greatly reduce the
\be9 production rate.  The resulting \be9 abundance
then is quite small, below observable levels
\cite{kusakabe2010}.

Bound-state effects are important in the context of minimal supersymmetry.
Substantial parameter space exists in which
the gravitino is the LSP and thus the dark matter,
while 
the next-to-lightest partner is the charged stau $\tilde\tau^\pm$, the scalar
partner to the tau lepton.
These models are probed by bound-state BBN, which imposes constraints
that are complementary to accelerator limits
\cite{kusakabe2007,bird2008,kkmy2008,kusakabe2008,cefos,pps2008,bjm2009,kt2010}.  
Indeed,
the solutions to the \li7 and possibly also the \li6 problem
which exist at $\tau_X \sim 1000$ sec
could be interpreted as support for these models.

To summarize, decaying dark matter scenarios
introduce a rich array of novel processes that
can alter light elements during and after BBN.
Moreover, such scenarios find well-motivated origin
in supersymmetric cosmologies.  Indeed, decaying-particle BBN
offers important constraints on supersymmetry. 
Furthermore, the \li7 and possibly also \li6 problems
can be solved in decaying particle scenarios 
which are realized in plausible minimal supersymmetric
scenarios.  This area is ripe for further theoretical, observational,
and experimental development.

\subsubsection{Changing Fundamental Constants}

Observations of multiple atomic transitions in
metals residing in 
high-redshift quasar absorption systems 
test
fundamental physics in environments at great
spacetime separations from our own;
for a review see ref.~\cite{uzan}.
Surprisingly, 
some data hint at variations in the fine-structure
constant at $z \sim 3$, showing
$\delta \alpha_{\rm EM}/\alpha_{\rm EM} \simeq - 0.5 \times 10^{-5}$
at the $\sim 5\sigma$ level,
while others
find results consistent with no variation.
Thus the observational situation is unsettled,
but intriguing.

Time variations in low-energy physics
can be accommodated 
and are even expected in the context of some unified theories
which in general predict stronger variations at earlier times.
Moreover, an underlying unified theory implies that
all Standard Model couplings and particle masses should vary, with definite
but model-dependent
interrelationships.

There is thus theoretical impetus, and some observational
motivation, to contemplate changes in fundamental constants
during BBN.  The change in light
elements depends on which parameters (couplings, masses)
change, and on the size of the perturbations
\cite{fs2003,flambaum2007,coc2007,berengut,coc2009}.
In general, there is large model-dependence
in quantifying the these variations, the links
among them, and their manifestation in nuclear properties
(masses, binding energies, cross sections).

An alternative approach is to turn the problem around
and to consider the BBN implications of variation in
nuclear physics parameters \cite{fs2002,dmitriev2004}.
Coc et al.~\cite{coc2007} 
systematically study the light-element response
to variations in 
$\alpha_{\rm EM}$, the electron mass $m_e$, the
neutron lifetime $\tau_n$, the
neutron-proton mass difference $m_n-m_p$, and the
deuteron binding energy $B_D$.
Of these, the most sensitive parameter is the
deuteron binding energy.
A change of $-0.075 \la \delta B_D/B_D \la -0.04$
lowers the \li7 abundance into concordance,
without perturbing \he4 or D/H beyond their observed error range.
Thus, unified models which predict changes of this order
can solve the lithium problem.

\subsubsection{Nonstandard Cosmologies}

The lithium problem 
could indicate nonstandard cosmology rather than
particle physics.
One recent such proposal is that cosmic acceleration could result from
large-scale inhomogeneities in the cosmic density.
Isotropy constraints can be satisfied 
if we occupy a privileged view from nearly the center of a spherically 
symmetric cosmic underdensity, which only returns to the 
cosmic mean at horizon-scale distances \cite{moffat}.
Such a scenario explains cosmic
acceleration
within General Relativity, and without invoking dark energy,
but must abandon the cosmological principle, instead requiring that
ours is a privileged  view of the universe. 

Esthetics aside, 
observations probe such an inhomogeneous cosmos.
BBN occurs differently in such a universe,
if the baryon-to-photon ratio varies along the inhomogeneity.
In particular, ref.~\cite{rc} emphasizes that
the observations of \li7 are made {\em locally}, at
low $z$, while D/H and the CMB are both
measured in the distant universe at high $z$.
If the local baryon-to-photon ratio $\eta_0$ is low by a factor $\sim 2$, then 
indeed one would expect local \li7 to fall below the WMAP
prediction, while D/H would agree.

This clever scenario however must face an array of observational tests.
Relaxed galaxy clusters probe the cosmic baryon-to-matter fraction,
and show variations $\la 8\%$ out to $z \sim 1$, far less than
the $\sim 50\%$ variation needed for lithium abundances \cite{hnv}.
Moreover, local measurements of D/H even more directly 
constrain the local $\eta$ measurement.
Because some stellar destruction may well have occurred, 
the local D/H sets a lower limit to the primordial abundance.
In the lower halo of our own Galaxy, 
${\rm D/H} = (2.31 \pm 0.24) \times 10^{-5}$.
This value is in good agreement with
the high-$z$ D/H measurements and the WMAP+BBN predictions;
it is therefore inconsistent with the low $\eta$ needed
for a low local \li7/H.


Other nonstandard cosmology scenarios have been proposed to
solve the lithium problem.
One suggests that a large fraction 
($\sim 1/3 - 1/2$) of baryons
were processed through the first generation of stars
(Population III) which ejected lithium-free matter
\cite{piau}.
Such models face grate difficulties
due to the substantial D/H depletion and \he3 production
which must also occur
\cite{foscv,vsof}.

\section{Discussion and Outlook}

\label{sect:outlook}

BBN has entered the age of precision cosmology.
This transition has brought triumph in the
spectacular agreement between high-redshift deuterium
and the BBN+WMAP prediction, and in the
WMAP confirmation of the longstanding BBN prediction
of nonbaryonic dark matter.
But new precision has raised new questions:
the measured primordial \li7 abundance 
falls persistently and significantly below BBN+WMAP predictions.
Moreover, there are controversial hints of a primordial
\li6 abundance orders of magnitude above the standard prediction.

As we have seen, disparate explanations for the lithium 
problem(s) remain viable.
Fortunately, most alternatives
are testable in the near future.
\begin{enumerate}

\item
{\em Astronomical observations.}
Recent indications of lithium depletion in extremely
metal-poor halo stars are tantalizing.
In the coming Great Survey era, we may expect many more
such stars to be identified, and the lithium trends
explored in large statistical samples.
These will require careful comparison with theory.
Observations of \li6 remain challenging, and
as yet it remains unclear what trends exist with metallicity.

Great insight would result from alternative measures of primordial lithium,
e.g., in the interstellar medium of metal-poor galaxies
nearby or at high redshift.

\item
{\em Nuclear experiments.}
The enormous effort of the nuclear community has
empirically pinned down nearly all nuclear inputs
to BBN.  Remaining are a few
known  or proposed resonances which would amplify
\be7 destruction.  These 
are within reach of present facilities, so that
the nuclear physics of standard BBN 
can and will be fully tested.

\item
{\em Collider and dark matter experiments.}
The LHC is operational and much of minimal supersymmetry
lies within its reach.  The discovery of supersymmetry
would revolutionize particle physics and cosmology,
and would transform decaying particle BBN scenarios 
into canonical early universe cosmology.  Alternatively,
if the LHC fails to find supersymmetry and/or finds 
surprises of some other kind, this will represent a paradigm
shift for all of particle physics
and particle cosmology, and BBN will lie at the heart of this transformation.

\end{enumerate}



\section{Disclosure Statement}

The author is unaware of any affiliations, memberships,
funding, or financial holdings that might
be perceived as affecting the objectivity of this review.

\section{Acknowledgments}

It is a pleasure to thank my collaborators
in primordial nucleosynthesis and closely related areas:  
Nachiketa Chakraborty, Richard Cyburt, 
John Ellis, Feng Luo, Keith Olive, Tijana Prodanovi\'{c},
Evan Skillman,
Vassilis Spanos, and Gary Steigman.
I am particularly grateful to Keith Olive for comments on
an earlier version of this paper.

\bibliographystyle{myarnuke}

\bibliography{lirefs}{}

\begin{thebibliography}{100}

\bibitem{wfh}
{Wagoner} RV, {Fowler} WA, {Hoyle} F,
\newblock \apj 148:3 (1967).

\bibitem{gary}
Steigman G,
\newblock \arnps 57:463 (2007), 0712.1100.

\bibitem{wmap1}
{Spergel} DN, {Verde} L, {Peiris} HV, {Komatsu} E, {Nolta} MR, et~al.,
\newblock \apjs 148:175 (2003), arXiv:astro-ph/0302209.

\bibitem{wmap7}
{Larson} D, {Dunkley} J, {Hinshaw} G, {Komatsu} E, {Nolta} MR, et~al.,
\newblock ArXiv e-prints  (2010), 1001.4635.

\bibitem{riess98}
{Riess} AG, {Filippenko} AV, {Challis} P, {Clocchiatti} A, {Diercks} A, et~al.,
\newblock \aj 116:1009 (1998), arXiv:astro-ph/9805201.

\bibitem{perlmutter99}
{Perlmutter} S, {Aldering} G, {Goldhaber} G, {Knop} RA, {Nugent} P, et~al.,
\newblock \apj 517:565 (1999), arXiv:astro-ph/9812133.

\bibitem{tonry03}
{Tonry} JL, {Schmidt} BP, {Barris} B, {Candia} P, {Challis} P, et~al.,
\newblock \apj 594:1 (2003), arXiv:astro-ph/0305008.

\bibitem{wood-vasey07}
{Wood-Vasey} WM, {Miknaitis} G, {Stubbs} CW, {Jha} S, {Riess} AG, et~al.,
\newblock \apj 666:694 (2007), arXiv:astro-ph/0701041.

\bibitem{st}
{Schramm} DN, {Turner} MS,
\newblock \rmp 70:303 (1998), arXiv:astro-ph/9706069.

\bibitem{cfo2002}
{Cyburt} RH, {Fields} BD, {Olive} KA,
\newblock Astropart.~Phys. 17:87 (2002), arXiv:astro-ph/0105397.

\bibitem{cfo2003}
Cyburt RH, Fields BD, Olive KA,
\newblock Phys.~Lett. B567:227 (2003), astro-ph/0302431.

\bibitem{cyburt}
Cyburt RH,
\newblock \prd 70:023505 (2004), astro-ph/0401091.

\bibitem{coc}
{Coc} A, {Vangioni-Flam} E, {Descouvemont} P, {Adahchour} A, {Angulo} C,
\newblock \apj 600:544 (2004), arXiv:astro-ph/0309480.

\bibitem{cuoco}
{Cuoco} A, {Iocco} F, {Mangano} G, {Miele} G, {Pisanti} O, {Serpico} PD,
\newblock International Journal of Modern Physics A 19:4431 (2004),
  arXiv:astro-ph/0307213.

\bibitem{pp2010}
Pospelov M, Pradler J,
\newblock \arnps 60:539 (2010), 1011.1054.

\bibitem{jp2009}
Jedamzik K, Pospelov M,
\newblock New J.Phys. 11:105028 (2009), 0906.2087.

\bibitem{iocco2009}
{Iocco} F, {Mangano} G, {Miele} G, {Pisanti} O, {Serpico} PD,
\newblock \physrep 472:1 (2009), 0809.0631.

\bibitem{kr}
{Krauss} LM, {Romanelli} P,
\newblock \apj 358:47 (1990).

\bibitem{skm}
{Smith} MS, {Kawano} LH, {Malaney} RA,
\newblock \apjs 85:219 (1993).

\bibitem{fiorentini}
{Fiorentini} G, {Lisi} E, {Sarkar} S, {Villante} FL,
\newblock \prd 58:063506 (1998), arXiv:astro-ph/9803177.

\bibitem{hata}
Hata N, Scherrer R, Steigman G, Thomas D, Walker T, et~al.,
\newblock \prl 75:3977 (1995), hep-ph/9505319.

\bibitem{nb}
{Nollett} KM, {Burles} S,
\newblock \prd 61:123505 (2000), arXiv:astro-ph/0001440.

\bibitem{cfo2001}
Cyburt RH, Fields BD, Olive KA,
\newblock New Astron. 6:215 (2001), astro-ph/0102179.

\bibitem{coc2003}
Coc A, Vangioni-Flam E, Descouvemont P, Adahchour A, Angulo C,
\newblock \apj 600:544 (2004), astro-ph/0309480.

\bibitem{descouvemont}
{Descouvemont} P, {Adahchour} A, {Angulo} C, {Coc} A, {Vangioni-Flam} E,
\newblock Atomic Data and Nuclear Data Tables 88:203 (2004),
  arXiv:astro-ph/0407101.

\bibitem{serpico}
{Serpico} PD, {Esposito} S, {Iocco} F, {Mangano} G, {Miele} G, {Pisanti} O,
\newblock \jcap 12:10 (2004), arXiv:astro-ph/0408076.

\bibitem{rishi}
{Khatri} R, {Sunyaev} RA,
\newblock ArXiv e-prints  (2010), 1009.3932.

\bibitem{tsof}
{Thomas} D, {Schramm} DN, {Olive} KA, {Fields} BD,
\newblock \apj 406:569 (1993), arXiv:astro-ph/9206002.

\bibitem{elisa99}
{Vangioni-Flam} E, {Casse} M, {Cayrel} R, {Audouze} J, {Spite} M, {Spite} F,
\newblock New Astron.~ 4:245 (1999), arXiv:astro-ph/9811327.

\bibitem{cfo2008}
{Cyburt} RH, {Fields} BD, {Olive} KA,
\newblock \jcap 11:12 (2008), 0808.2818.

\bibitem{ceflos2009}
{Cyburt} RH, {Ellis} J, {Fields} BD, {Luo} F, {Olive} KA, {Spanos} VC,
\newblock \jcap 10:21 (2009), 0907.5003.

\bibitem{ceflos2010}
{Cyburt} RH, {Ellis} J, {Fields} BD, {Luo} F, {Olive} KA, {Spanos} VC,
\newblock \jcap 10:32 (2010), 1007.4173.

\bibitem{bt98}
{Burles} S, {Tytler} D,
\newblock \apj 507:732 (1998), arXiv:astro-ph/9712109.

\bibitem{bt98b}
{Burles} S, {Tytler} D,
\newblock \apj 499:699 (1998), arXiv:astro-ph/9712108.

\bibitem{omeara2001}
{O'Meara} JM, {Tytler} D, {Kirkman} D, {Suzuki} N, {Prochaska} JX, et~al.,
\newblock \apj 552:718 (2001), arXiv:astro-ph/0011179.

\bibitem{kirkman}
{Kirkman} D, {Tytler} D, {Suzuki} N, {O'Meara} JM, {Lubin} D,
\newblock \apjs 149:1 (2003), arXiv:astro-ph/0302006.

\bibitem{omeara2006}
{O'Meara} JM, {Burles} S, {Prochaska} JX, {Prochter} GE, {Bernstein} RA,
  {Burgess} KM,
\newblock \apjl 649:L61 (2006), arXiv:astro-ph/0608302.

\bibitem{pettini2008}
{Pettini} M, {Zych} BJ, {Murphy} MT, {Lewis} A, {Steidel} CC,
\newblock Mon.~Not.~Roy.~Astr.~Soc.~ 391:1499 (2008), 0805.0594.

\bibitem{os2001}
{Olive} KA, {Skillman} ED,
\newblock New Astron.~ 6:119 (2001).

\bibitem{os2004}
{Olive} KA, {Skillman} ED,
\newblock \apj 617:29 (2004), arXiv:astro-ph/0405588.

\bibitem{peimbert}
{Peimbert} A, {Peimbert} M, {Luridiana} V,
\newblock \apj 565:668 (2002), arXiv:astro-ph/0107189.

\bibitem{izotov}
{Izotov} YI, {Thuan} TX,
\newblock \apj 500:188 (1998).

\bibitem{thuan}
{Izotov} YI, {Thuan} TX,
\newblock \apj 602:200 (2004), arXiv:astro-ph/0310421.

\bibitem{bania}
{Bania} TM, {Rood} RT, {Balser} DS,
\newblock Space Sci.~Rev.~ 130:53 (2007).

\bibitem{vofc}
{Vangioni-Flam} E, {Olive} KA, {Fields} BD, {Cass{\'e}} M,
\newblock \apj 585:611 (2003), arXiv:astro-ph/0207583.

\bibitem{asplund}
{Asplund} M, {Lambert} DL, {Nissen} PE, {Primas} F, {Smith} VV,
\newblock \apj 644:229 (2006), arXiv:astro-ph/0510636.

\bibitem{spite}
{Spite} F, {Spite} M,
\newblock \aap 115:357 (1982).

\bibitem{rnb}
{Ryan} SG, {Norris} JE, {Beers} TC,
\newblock \apj 523:654 (1999), arXiv:astro-ph/9903059.

\bibitem{bonifacio}
{Bonifacio} P, {Molaro} P, {Sivarani} T, {Cayrel} R, {Spite} M, et~al.,
\newblock \aap 462:851 (2007), arXiv:astro-ph/0610245.

\bibitem{korn}
{Korn} AJ, {Grundahl} F, {Richard} O, {Barklem} PS, {Mashonkina} L, et~al.,
\newblock \nat 442:657 (2006), arXiv:astro-ph/0608201.

\bibitem{sbordone}
{Sbordone} L, {Bonifacio} P, {Caffau} E, {Ludwig} H, {Behara} NT, et~al.,
\newblock \aap 522:A26+ (2010), 1003.4510.

\bibitem{aoki}
{Aoki} W, {Barklem} PS, {Beers} TC, {Christlieb} N, {Inoue} S, et~al.,
\newblock \apj 698:1803 (2009), 0904.1448.

\bibitem{hosford09}
{Hosford} A, {Ryan} SG, {Garc{\'{\i}}a P{\'e}rez} AE, {Norris} JE, {Olive} KA,
\newblock \aap 493:601 (2009), 0811.2506.

\bibitem{hosford10}
{Hosford} A, {Garc{\'{\i}}a P{\'e}rez} AE, {Collet} R, {Ryan} SG, {Norris} JE,
  {Olive} KA,
\newblock \aap 511:A47+ (2010), 1004.0863.

\bibitem{melendez}
{Mel{\'e}ndez} J, {Casagrande} L, {Ram{\'{\i}}rez} I, {Asplund} M, {Schuster}
  WJ,
\newblock \aap 515:L3+ (2010), 1005.2944.

\bibitem{g-h}
{Gonz{\'a}lez Hern{\'a}ndez} JI, {Bonifacio} P, {Caffau} E, {Steffen} M,
  {Ludwig} H, et~al.,
\newblock \aap 505:L13 (2009), 0909.0983.

\bibitem{rbofn}
{Ryan} SG, {Beers} TC, {Olive} KA, {Fields} BD, {Norris} JE,
\newblock \apjl 530:L57 (2000), arXiv:astro-ph/9905211.

\bibitem{monaco}
{Monaco} L, {Bonifacio} P, {Sbordone} L, {Villanova} S, {Pancino} E,
\newblock \aap 519:L3+ (2010), 1008.1817.

\bibitem{cayrel2007}
{Cayrel} R, {Steffen} M, {Chand} H, {Bonifacio} P, {Spite} M, et~al.,
\newblock \aap 473:L37 (2007), 0708.3819.

\bibitem{hd}
{Hu} W, {Dodelson} S,
\newblock \araa 40:171 (2002), arXiv:astro-ph/0110414.

\bibitem{ytsso}
{Yang} J, {Turner} MS, {Schramm} DN, {Steigman} G, {Olive} KA,
\newblock \apj 281:493 (1984).

\bibitem{wssof}
{Walker} TP, {Steigman} G, {Kang} H, {Schramm} DM, {Olive} KA,
\newblock \apj 376:51 (1991).

\bibitem{wmap5}
{Komatsu} E, {Dunkley} J, {Nolta} MR, {Bennett} CL, {Gold} B, et~al.,
\newblock \apjs 180:330 (2009), 0803.0547.

\bibitem{fov}
{Fields} BD, {Olive} KA, {Vangioni-Flam} E,
\newblock \apj 623:1083 (2005), arXiv:astro-ph/0411728.

\bibitem{mr}
{Mel{\'e}ndez} J, {Ram{\'{\i}}rez} I,
\newblock \apjl 615:L33 (2004), arXiv:astro-ph/0409383.

\bibitem{casagrande}
{Casagrande} L, {Ram{\'{\i}}rez} I, {Mel{\'e}ndez} J, {Bessell} M, {Asplund} M,
\newblock \aap 512:A54+ (2010), 1001.3142.

\bibitem{wallerstein}
{Wallerstein} G, {Conti} PS,
\newblock \araa 7:99 (1969).

\bibitem{spites}
{Spite} M, {Spite} F,
\newblock \araa 23:225 (1985).

\bibitem{gustafsson}
{Gustafsson} B,
\newblock \araa 27:701 (1989).

\bibitem{pinsonneault}
{Pinsonneault} M,
\newblock \araa 35:557 (1997).

\bibitem{bodenheimer}
{Bodenheimer} P,
\newblock \apj 142:451 (1965).

\bibitem{iben}
{Iben} Jr. I,
\newblock \apj 147:624 (1967).

\bibitem{pdd}
{Pinsonneault} MH, {Deliyannis} CP, {Demarque} P,
\newblock \apjs 78:179 (1992).

\bibitem{charbonnel}
{Talon} S, {Charbonnel} C,
\newblock \aap 418:1051 (2004), arXiv:astro-ph/0401474.

\bibitem{richard}
{Richard} O, {Michaud} G, {Richer} J,
\newblock \apj 619:538 (2005), arXiv:astro-ph/0409672.

\bibitem{gr}
{Greenstein} JL, {Richardson} RS,
\newblock \apj 113:536 (1951).

\bibitem{bs}
{Brown} L, {Schramm} DN,
\newblock \apjl 329:L103 (1988).

\bibitem{mf}
{Malaney} RA, {Fowler} WA,
\newblock \apjl 345:L5 (1989).

\bibitem{tsommf}
{Thomas} D, {Schramm} DN, {Olive} KA, {Mathews} GJ, {Meyer} BS, {Fields} BD,
\newblock \apj 430:291 (1994), arXiv:astro-ph/9308026.

\bibitem{iocco2007}
{Iocco} F, {Mangano} G, {Miele} G, {Pisanti} O, {Serpico} PD,
\newblock \prd 75:087304 (2007), arXiv:astro-ph/0702090.

\bibitem{vcc}
{Vangioni-Flam} E, {Coc} A, {Cass{\'e}} M,
\newblock \aap 360:15 (2000), arXiv:astro-ph/0002248.

\bibitem{boyd}
{Boyd} RN, {Brune} CR, {Fuller} GM, {Smith} CJ,
\newblock \prd 82:105005 (2010), 1008.0848.

\bibitem{chfo}
{Chakraborty} N, {Fields} BD, {Olive} KA,
\newblock ArXiv e-prints  (2010), 1011.0722.

\bibitem{cd}
{Cyburt} RH, {Davids} B,
\newblock \prc 78:064614 (2008), 0809.3240.

\bibitem{cfo2004}
{Cyburt} RH, {Fields} BD, {Olive} KA,
\newblock \prd 69:123519 (2004), arXiv:astro-ph/0312629.

\bibitem{dicus}
Dicus DA, Kolb EW, Gleeson AM, Sudarshan ECG, Teplitz VL, Turner MS,
\newblock Phys. Rev. D 26:2694 (1982).

\bibitem{seckel}
{Seckel} D,
\newblock ArXiv High Energy Physics - Phenomenology e-prints  (1993),
  arXiv:hep-ph/9305311.

\bibitem{dolgov}
{Dolgov} AD, {Fukugita} M,
\newblock \prd 46:5378 (1992).

\bibitem{kernan}
{Kernan} PJ, {Krauss} LM,
\newblock Physical Review Letters 72:3309 (1994), arXiv:astro-ph/9402010.

\bibitem{dodelson}
{Dodelson} S, {Turner} MS,
\newblock \prd 46:3372 (1992).

\bibitem{hannestad}
{Hannestad} S, {Madsen} J,
\newblock \prd 52:1764 (1995), arXiv:astro-ph/9506015.

\bibitem{lopez}
Lopez RE, Turner MS,
\newblock Phys. Rev. D 59:103502 (1999).

\bibitem{esposito}
{Esposito} S, {Mangano} G, {Miele} G, {Pisanti} O,
\newblock Nuclear Physics B 540:3 (1999), arXiv:astro-ph/9808196.

\bibitem{sf}
{Smith} CJ, {Fuller} GM,
\newblock \prd 81:065027 (2010), 0905.2781.

\bibitem{vkk}
{Voronchev} VT, {Nakao} Y, {Nakamura} M,
\newblock \apj 725:242 (2010).

\bibitem{itoh}
{Itoh} N, {Nishikawa} A, {Nozawa} S, {Kohyama} Y,
\newblock \apj 488:507 (1997).

\bibitem{angulo}
{Angulo} C, {Casarejos} E, {Couder} M, {Demaret} P, {Leleux} P, et~al.,
\newblock \apjl 630:L105 (2005), arXiv:astro-ph/0508454.

\bibitem{cp}
{Cyburt} RH, {Pospelov} M,
\newblock ArXiv e-prints  (2009), 0906.4373.

\bibitem{chakraborty}
{Chakraborty} N, {Fields} BD, {Olive} KA,
\newblock ArXiv e-prints  (2010), 1011.0722.

\bibitem{hoyle}
{Hoyle} F,
\newblock \apjs 1:121 (1954).

\bibitem{feng}
{Feng} JL,
\newblock \araa 48:495 (2010), 1003.0904.

\bibitem{hooper}
{Hooper} D, {Baltz} EA,
\newblock Annual Review of Nuclear and Particle Science 58:293 (2008),
  0802.0702.

\bibitem{gaitskell}
{Gaitskell} RJ,
\newblock Annual Review of Nuclear and Particle Science 54:315 (2004).

\bibitem{ekn84}
{Ellis} J, {Kim} JE, {Nanopoulos} DV,
\newblock Physics Letters B 145:181 (1984).

\bibitem{lindley85}
{Lindley} D,
\newblock \apj 294:1 (1985).

\bibitem{ens85}
{Ellis} J, {Nanopoulos} DV, {Sarkar} S,
\newblock Nuclear Physics B 259:175 (1985).

\bibitem{jss85}
{Juszkiewicz} R, {Silk} J, {Stebbins} A,
\newblock Physics Letters B 158:463 (1985).

\bibitem{ks87}
{Kawasaki} M, {Sato} K,
\newblock Physics Letters B 189:23 (1987).

\bibitem{reno88}
{Reno} MH, {Seckel} D,
\newblock \prd 37:3441 (1988).

\bibitem{scherrer88}
{Scherrer} RJ, {Turner} MS,
\newblock \apj 331:19 (1988).

\bibitem{dehs89}
{Dimopoulos} S, {Esmailzadeh} R, {Hall} LJ, {Starkman} GD,
\newblock Nuclear Physics B 311:699 (1989).

\bibitem{eglns92}
{Ellis} J, {Gelmini} GB, {Lopez} JL, {Nanopoulos} DV, {Sarkar} S,
\newblock Nuclear Physics B 373:399 (1992).

\bibitem{moroi93}
{Moroi} T, {Murayama} H, {Yamaguchi} M,
\newblock Physics Letters B 303:289 (1993).

\bibitem{jedamzik2006}
{Jedamzik} K,
\newblock \prd 74:103509 (2006), arXiv:hep-ph/0604251.

\bibitem{cefo}
{Cyburt} RH, {Ellis} J, {Fields} BD, {Olive} KA,
\newblock \prd 67:103521 (2003), arXiv:astro-ph/0211258.

\bibitem{kkm2005}
{Kawasaki} M, {Kohri} K, {Moroi} T,
\newblock \prd 71:083502 (2005), arXiv:astro-ph/0408426.

\bibitem{pospelov2007}
{Pospelov} M,
\newblock Physical Review Letters 98:231301 (2007), arXiv:hep-ph/0605215.

\bibitem{jed2004li6}
{Jedamzik} K,
\newblock \prd 70:083510 (2004), arXiv:astro-ph/0405583.

\bibitem{jedamzik2004}
{Jedamzik} K,
\newblock \prd 70:063524 (2004), arXiv:astro-ph/0402344.

\bibitem{jcrr2006}
{Jedamzik} K, {Choi} K, {Roszkowski} L, {Ruiz de Austri} R,
\newblock \jcap 7:7 (2006), arXiv:hep-ph/0512044.

\bibitem{kusakabe2007}
{Kusakabe} M, {Kajino} T, {Boyd} RN, {Yoshida} T, {Mathews} GJ,
\newblock \prd 76:121302 (2007), 0711.3854.

\bibitem{pps2008}
{Pospelov} M, {Pradler} J, {Steffen} FD,
\newblock \jcap 11:20 (2008), 0807.4287.

\bibitem{jedamzik2008}
{Jedamzik} K,
\newblock \prd 77:063524 (2008), 0707.2070.

\bibitem{bjm2009}
{Bailly} S, {Jedamzik} K, {Moultaka} G,
\newblock \prd 80:063509 (2009), 0812.0788.

\bibitem{peskin}
{Peskin} ME,
\newblock {\em {Supersymmetry in Elementary Particle Physics}} (World
  Scientific, 2008), pp. 609--+.

\bibitem{olive}
{Olive} KA,
\newblock {Course 5: Introduction to Supersymmetry: Astrophysical and
  Phenomenological Constraints},
\newblock in {\em The Primordial Universe}, pp. 221--+, 2000,
  arXiv:hep-ph/9911307.

\bibitem{martin}
{Martin} SP,
\newblock {A Supersymmetry Primer},
\newblock in {\em Perspectives on Supersymmetry}, pp. 1--+, 1998,
  arXiv:hep-ph/9709356.

\bibitem{ehnos}
{Ellis} J, {Hagelin} JS, {Nanopoulos} DV, {Olive} K, {Srednicki} M,
\newblock Nuclear Physics B 238:453 (1984).

\bibitem{eoss}
{Ellis} J, {Olive} KA, {Santoso} Y, {Spanos} VC,
\newblock Physics Letters B 565:176 (2003), arXiv:hep-ph/0303043.

\bibitem{kamimura}
{Kamimura} M, {Kino} Y, {Hiyama} E,
\newblock Progress of Theoretical Physics 121:1059 (2009), 0809.4772.

\bibitem{kusakabe2010}
{Kusakabe} M, {Kajino} T, {Yoshida} T, {Mathews} GJ,
\newblock \prd 81:083521 (2010), 1001.1410.

\bibitem{bird2008}
{Bird} C, {Koopmans} K, {Pospelov} M,
\newblock \prd 78:083010 (2008), arXiv:hep-ph/0703096.

\bibitem{kusakabe2008}
{Kusakabe} M, {Kajino} T, {Boyd} RN, {Yoshida} T, {Mathews} GJ,
\newblock \apj 680:846 (2008), 0711.3858.

\bibitem{cefos}
{Cyburt} RH, {Ellis} J, {Fields} BD, {Olive} KA, {Spanos} VC,
\newblock \jcap 11:14 (2006), arXiv:astro-ph/0608562.

\bibitem{kkmy2008}
{Kawasaki} M, {Kohri} K, {Moroi} T, {Yotsuyanagi} A,
\newblock \prd 78:065011 (2008), 0804.3745.

\bibitem{kt2010}
{Kohri} K, {Takahashi} T,
\newblock Physics Letters B 682:337 (2010), 0909.4610.

\bibitem{uzan}
{Uzan} J,
\newblock Reviews of Modern Physics 75:403 (2003), arXiv:hep-ph/0205340.

\bibitem{fs2003}
{Flambaum} VV, {Shuryak} EV,
\newblock \prd 67:083507 (2003), arXiv:hep-ph/0212403.

\bibitem{flambaum2007}
{Flambaum} VV, {Wiringa} RB,
\newblock \prc 76:054002 (2007), 0709.0077.

\bibitem{coc2007}
{Coc} A, {Nunes} NJ, {Olive} KA, {Uzan} J, {Vangioni} E,
\newblock \prd 76:023511 (2007), arXiv:astro-ph/0610733.

\bibitem{berengut}
{Berengut} JC, {Flambaum} VV, {Dmitriev} VF,
\newblock Physics Letters B 683:114 (2010), 0907.2288.

\bibitem{coc2009}
{Coc} A, {Olive} KA, {Uzan} J, {Vangioni} E,
\newblock \prd 79:103512 (2009), 0811.1845.

\bibitem{fs2002}
{Flambaum} VV, {Shuryak} EV,
\newblock \prd 65:103503 (2002), arXiv:hep-ph/0201303.

\bibitem{dmitriev2004}
{Dmitriev} VF, {Flambaum} VV, {Webb} JK,
\newblock \prd 69:063506 (2004), arXiv:astro-ph/0310892.

\bibitem{moffat}
{Moffat} JW,
\newblock \jcap 5:1 (2006), arXiv:astro-ph/0505326.

\bibitem{rc}
{Regis} M, {Clarkson} C,
\newblock ArXiv e-prints  (2010), 1003.1043.

\bibitem{hnv}
{Holder} GP, {Nollett} KM, {van Engelen} A,
\newblock \apj 716:907 (2010), 0907.3919.

\bibitem{piau}
{Piau} L, {Beers} TC, {Balsara} DS, {Sivarani} T, {Truran} JW, {Ferguson} JW,
\newblock \apj 653:300 (2006), arXiv:astro-ph/0603553.

\bibitem{foscv}
{Fields} BD, {Olive} KA, {Silk} J, {Cass{\'e}} M, {Vangioni-Flam} E,
\newblock \apj 563:653 (2001), arXiv:astro-ph/0107389.

\bibitem{vsof}
{Vangioni} E, {Silk} J, {Olive} KA, {Fields} BD,
\newblock ArXiv e-prints  (2010), 1010.5726.

\end{thebibliography}

\end{document}